\documentclass[sigconf]{acmart}
\usepackage{amsmath,amsfonts}
\usepackage{algorithmic}
\usepackage{graphicx}
\usepackage{textcomp}
\usepackage{subcaption}
\usepackage{xcolor}
\usepackage{fancyhdr}
\usepackage{hyperref}
\usepackage{xspace}
\usepackage{enumitem}
\usepackage{booktabs}
\usepackage{colortbl}
\usepackage{multirow}
\usepackage{pifont}
\usepackage{makecell}
\usepackage{cleveref}
\usepackage{soul,color}
\usepackage{acmart-taps}

\newcommand{\methodname}{\texttt{Ecco}\xspace}
\newcommand{\x}{$\times$\xspace}

\copyrightyear{2025}
\acmYear{2025}
\setcopyright{cc}
\setcctype{by-nc}
\acmConference[ISCA '25]{Proceedings of the 52nd Annual International Symposium on Computer Architecture}{June 21--25, 2025}{Tokyo, Japan}
\acmBooktitle{Proceedings of the 52nd Annual International Symposium on Computer Architecture (ISCA '25), June 21--25, 2025, Tokyo, Japan}\acmDOI{10.1145/3695053.3731024}
\acmISBN{979-8-4007-1261-6/2025/06}
\begin{document}
\title{Ecco: Improving Memory Bandwidth and Capacity for LLMs via \texorpdfstring{\underline{E}}{E}ntropy-aware \texorpdfstring{\underline{C}}{C}ache \texorpdfstring{\underline{Co}}{Co}mpression}

% %%%%%%%%%%%%%%%%%%%%%%%%%%%%%%%%%%%%%%%%
% %%%%%%%%%%%%%% -- AUTHORLIST -- %%%%%%%%%%%%%%%
% \author{
% Feng Cheng\textsuperscript{1},
% Cong Guo\textsuperscript{1},
% Chiyue Wei\textsuperscript{1},
% Junyao Zhang\textsuperscript{1},
% Changchun Zhou\textsuperscript{1}, \\
% Edward Hanson\textsuperscript{2},
% Jiaqi Zhang\textsuperscript{2},
% Xiaoxiao Liu\textsuperscript{2},
% Hai ``Helen'' Li\textsuperscript{1},
% Yiran Chen\textsuperscript{1}
% }

% \affiliation{
% \normalsize{
% \institution{\textsuperscript{1}Duke University,
% \textsuperscript{2}Advanced Micro Devices, Inc.}
% \city{}
% \country{}}
% }

\author{Feng Cheng}
\affiliation{
  \institution{Duke University}
  \city{Durham}
  \country{USA}
}
\email{feng.cheng@duke.edu}

\author{Cong Guo}
\authornote{Cong Guo is the corresponding author of this paper.}
\affiliation{%
  \institution{Duke University}
  \city{Durham}
  \country{USA}
}
\email{cong.guo@duke.edu}

\author{Chiyue Wei}
\affiliation{
  \institution{Duke University}
  \city{Durham}
  \country{USA}
}
\email{chiyue.wei@duke.edu}

\author{Junyao Zhang}
\affiliation{
  \institution{Duke University}
  \city{Durham}
  \country{USA}
}
\email{jz420@duke.edu}

\author{Changchun Zhou}
\affiliation{
  \institution{Duke University}
  \city{Durham}
  \country{USA}
}
\email{changchun.zhou@duke.edu}

\author{Edward Hanson}
\affiliation{
  \institution{Advanced Micro Devices, Inc.}
  \city{Santa Clara}
  \country{USA}
}
\email{edward.hanson@amd.com}

\author{Jiaqi Zhang}
\affiliation{
  \institution{Advanced Micro Devices, Inc.}
  \city{Santa Clara}
  \country{USA}
}
\email{jiaqi.zhang@amd.com}

\author{Xiaoxiao Liu}
\affiliation{
  \institution{Advanced Micro Devices, Inc.}
  \city{Santa Clara}
  \country{USA}
}
\email{xiaoxiao.liu@amd.com}

\author{Hai ``Helen'' Li}
\affiliation{
  \institution{Duke University}
  \city{Durham}
  \country{USA}
}
\email{hai.li@duke.edu}

\author{Yiran Chen}
\affiliation{
  \institution{Duke University}
  \city{Durham}
  \country{USA}
}
\email{yiran.chen@duke.edu}

\renewcommand{\shortauthors}{Cheng, et al.}

%%%%%%%%%%%%%%%%%%%%%%%%%%%%%%%%%%%%%%%%

\begin{abstract}

Large language models (LLMs) have demonstrated transformative capabilities across diverse artificial intelligence applications, yet their deployment is hindered by substantial memory and computational demands, especially in resource-constrained environments. Quantization techniques have emerged as a critical solution, reducing data precision to enhance memory and computational efficiency. 
However, existing methods often suffer from high runtime overheads and potential accuracy degradation. 
To address these challenges, we propose \textbf{\textit{Ecco}}, an entropy-based cache compression technique tailored for LLMs. 
Ecco combines group-wise and non-uniform quantization with pre-defined shared k-means patterns and Huffman coding to exploit the inherent entropy characteristics of LLM cache data. 
Recognizing the inefficiencies of traditional Huffman coding in terms of parallelism and latency, we introduce a novel parallel Huffman-based decoding process with a multi-stage pipeline design, reducing latency by two orders of magnitude and achieving throughput comparable to GPU L2 caches. 
Comprehensive evaluations demonstrate that Ecco achieves an up to 2.9$\times$ and 1.9$\times$ speedup over the state-of-the-art AWQ and SmoothQuant framework, 2.4$\times$ over the Olive accelerator, all while increasing memory capacity by nearly 4$\times$ and maintaining state-of-the-art LLM accuracy. 
These results underscore the effectiveness of our entropy-based cache compression in enhancing LLM performance and efficiency, paving the way for more deployable large-scale AI models.

\end{abstract}


% %%%%%%%%%%%%%%%%%%%%%%%%%%%%%%%%%%%%%%%%
% %%%%%%%%%%%%%% -- CCSCONCEPTS -- %%%%%%%%%%%%%%%

%%%%%%%%%%%%%%%%%%%%%%%%%%%%%%%%%%%%%%%%

\maketitle

%%%%%%%%%%%%%%%%%%%%%%%%%%%%%%%%%%%%%%%%
%%%%%%%% -- PAPER CONTENT STARTS -- %%%%%%%%%
\section{Introduction}

Large language models (LLMs) \cite{brown2020language,touvron2023llama,achiam2023gpt} have revolutionized artificial intelligence by demonstrating remarkable capabilities across diverse tasks, including natural language understanding \cite{llm_nlp}, generation \cite{llm_gen}, complex reasoning \cite{llm_reasoning}, and problem-solving \cite{llm_solve}. 
These advancements have made LLMs indispensable in modern AI applications. 
However, as LLMs scale, they introduce substantial memory and computational overheads that challenge their deployment efficiency, particularly in resource-constrained environments \cite{shoeybi2019megatron,kaplan2020scaling}.

LLMs typically operate in two distinct phases during inference: prefill and decode \cite{touvron2023llama,brown2020language,achiam2023gpt}. 
The memory-bound nature of LLMs is most pronounced due to these mechanisms \cite{yuan2024llm}. 
Firstly, the auto-regressive property requires the model to maintain and update a substantial amount of intermediate state information, 
the key-value (KV) cache, for each generated token. 
This results in extensive memory capacity demands, accounting for up to 72.7\% (34.4 GB) of total memory usage (47.3 GB) in an LLaMA-7B model with a 2K sequence length and a batch size of 32 \cite{yuan2024llm}. 
Secondly, the decoder-only architecture employs per-token inference with small-batch general matrix multiplication (GEMM) operations, leading to serious memory-bound and computational underutilization, thereby creating a new memory wall in the era of LLMs \cite{vllm}. 
This memory-intensive behavior is exacerbated as model sizes continue to grow, with state-of-the-art (SOTA) LLMs often containing billions of parameters \cite{touvron2023llama,achiam2023gpt}. 
Consequently, the performance of these models is frequently constrained by memory bandwidth and capacity rather than computational power, making memory management and optimization critical factors in enhancing LLM inference efficiency.

\begin{figure}
    \centering
    \includegraphics[width=\linewidth]{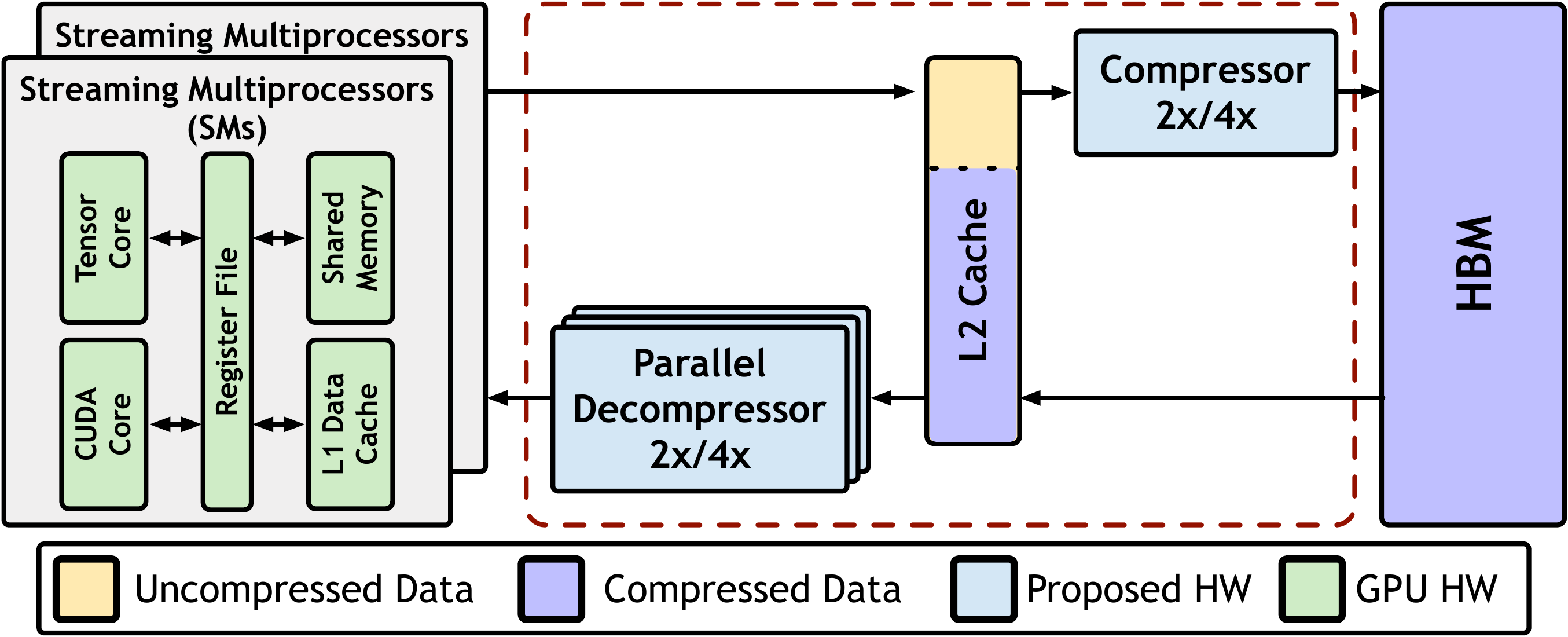}
    \caption{\methodname Overview: A high-throughput parallel compressor and decompressor are integrated with the L2 cache
    to address memory bandwidth and capacity constraints in LLMs. Data is compressed prior to transmission to high-bandwidth memory (HBM) and decompressed upon retrieval for use by streaming multiprocessors (SMs).}
    \label{fig:system_overview}
    \Description{}
\end{figure}

Model compression \cite{guo2020accelerating, kim2023squeezellm, lin2024awq, hooper2024kvquant, guo2025survey, guo2024accelerating, zhang2024dstc, dettmers2024qlora} has emerged as a promising solution to address these memory challenges in LLMs. 
Beyond reducing model size, compression techniques can significantly accelerate LLM inference by alleviating memory bandwidth constraints. 
Quantization \cite{jacob2018quantization}, a key compression technique, has been particularly effective in this domain \cite{hooper2024kvquant,lin2024awq,lin2024qserve, llmsurvey}. 
Recent SOTA studies, such as AWQ \cite{lin2024awq}, have achieved up to a 2-3$\times$ reduction in memory footprint, resulting in a 1.5$\times$ speedup in inference time, even with the original FP16 computation. 
This direct correlation between memory efficiency and computational performance has sparked unprecedented research interest in memory compression, driven by the pressing need to make LLMs more deployable across diverse hardware platforms.

However, existing compression methods for LLMs \cite{hooper2024kvquant,lin2024awq,lin2024qserve} face significant challenges in achieving optimal performance. 
The first notable issue is the high overhead associated with compression processes at runtime. 
Model compression requires frequent processing during inference, potentially negating the speed gains it offers. 
Moreover, aggressive compression techniques that significantly reduce model size may inadvertently compromise the accuracy of LLMs. 
Balancing memory efficiency, computational speedup, and model fidelity presents a formidable obstacle in developing effective compression methods for LLMs. 
Consequently, innovative approaches are urgently needed to address these challenges and bridge the gap between theoretical compression benefits and practical performance improvements.

Fortunately, these challenges can be tackled from a new computer architecture perspective by leveraging cache-level compression, as illustrated in \Cref{fig:system_overview}. 
Cache compression \cite{jain2018gist,pekhimenko2012base, alameldeen2004adaptive} offers a promising approach to unlock the latent performance potential of LLMs and has been widely adopted by commercial accelerators \cite{a100,amd2023mi300x}. 
For instance, the A100 GPU incorporates a novel cache compression mechanism \cite{a100} using a lossless compression method. 
This approach offers several distinct advantages over traditional model compression techniques. 
Firstly, cache compression typically incurs low overhead, as it operates at the cache level rather than being embedded into the model architecture or computation kernels. 
Secondly, leveraging cache compression allows for the design of more effective compression methods supported by hardware with minimal impact on model performance. 
This is crucial for maintaining the high quality of LLM outputs while efficiently addressing memory constraints.

In this study, we propose an innovative \underline{E}ntropy-based \underline{C}ache \underline{Co}mpression technique, termed \textbf{\textit{Ecco}}, tailored for LLMs. 
As shown in \Cref{fig:system_overview}, our approach introduces a novel compression method that combines group-wise and non-uniform quantization, utilizing pre-defined shared k-means \cite{macqueen1967some} patterns along with the Huffman coding \cite{huffman1952method} scheme. 
This hybrid technique effectively extracts and exploits the entropy characteristics inherent in LLM cache data. 
However, these methods present significant challenges for efficient implementation due to their irregular nature. 
For example, Huffman coding is a variable-length compression method, which limits its parallelism and efficiency \cite{huffman1952method,han2016eie,han2015deep}.

To overcome these challenges and enhance the throughput of our compression method, we propose a novel parallel Huffman-based decoding process with a multi-stage pipeline design. 
This approach reduces latency by two orders of magnitude compared to traditional sequential Huffman encoding methods \cite{huffman1952method} and achieves throughput comparable to the L2 cache in GPUs \cite{a100}. 
Additionally, we present an encoder design that efficiently matches the decoder's capabilities while ensuring practical implementation feasibility. 
Our method seamlessly integrates with traditional cache architectures, leveraging existing structures while introducing minimal overhead. 
Notably, our system supports runtime encoding and decoding, enabling dynamic compression and decompression of cache data as needed.

This adaptive approach allows us to achieve both high bandwidth utilization and substantially increased memory capacity. Our comprehensive evaluation demonstrates significant improvements, with our method achieving a speedup of up to 2.9$\times$ over the SOTA quantization framework AWQ \cite{lin2024awq} and 2.4$\times$ over the quantization accelerator Olive \cite{olive}, along with a memory capacity enlargement of nearly 4$\times$, while maintaining SOTA LLM accuracy. 
These results underscore the effectiveness of our entropy-based cache compression in enhancing LLM performance and efficiency.

We make the following contributions in this paper.
\aptLtoX[graphic=no,type=html]{\begin{itemize}%[leftmargin=1.2em]
    \item We introduce a novel compression paradigm, \textbf{\textit{Ecco}}, that combines group-wise and non-uniform quantization with Huffman coding to exploit entropy characteristics in LLM cache data, significantly enhancing memory efficiency.
    \item We design a parallel Huffman-based decoding process with a multi-stage pipeline, reducing latency by two orders of magnitude and achieving throughput comparable to GPU L2 caches, thereby overcoming the limitations of traditional Huffman coding.
    \item We develop an efficient encoder that complements our decoder design, enabling seamless integration with existing cache architectures and supporting runtime encoding and decoding for dynamic cache data management.
    \item Through comprehensive evaluations, we demonstrate that \textbf{\textit{Ecco}} achieves up to 2.9$\times$ speedup over AWQ framework and 2.4$\times$ over the Olive accelerator while increasing memory capacity by nearly 4$\times$ and maintaining SOTA LLM accuracy.
\end{itemize}}{\setlist{nolistsep}
\begin{itemize}[leftmargin=1.2em]
    \item We introduce a novel compression paradigm, \textbf{\textit{Ecco}}, that combines group-wise and non-uniform quantization with Huffman coding to exploit entropy characteristics in LLM cache data, significantly enhancing memory efficiency.
    \item We design a parallel Huffman-based decoding process with a multi-stage pipeline, reducing latency by two orders of magnitude and achieving throughput comparable to GPU L2 caches, thereby overcoming the limitations of traditional Huffman coding.
    \item We develop an efficient encoder that complements our decoder design, enabling seamless integration with existing cache architectures and supporting runtime encoding and decoding for dynamic cache data management.
    \item Through comprehensive evaluations, we demonstrate that \textbf{\textit{Ecco}} achieves up to 2.9$\times$ speedup over AWQ framework and 2.4$\times$ over the Olive accelerator while increasing memory capacity by nearly 4$\times$ and maintaining SOTA LLM accuracy.
\end{itemize}}

\section{Background and Motivation}

\subsection{LLM Inference}

Large language models (LLMs) have revolutionized natural language processing tasks by adopting the Transformer architecture \cite{vaswani2023attentionneed}. 
The Transformer consists of two primary components: the multi-head attention mechanism and the feed-forward network (FFN). The multi-head attention can be formulated as:
\begin{equation}
\text{MultiHead}(X) = \text{Concat}(\text{head}_1, \dots, \text{head}_h)W^O,
\end{equation}
where each head is computed as:
\begin{equation}
\text{head}_i = \text{Attention}(XW_i^Q, XW_i^K, XW_i^V),
\end{equation}
and the attention function is defined as:
\begin{equation}
\text{Attention}(Q, K, V) = \text{softmax}\left(\frac{QK^\top}{\sqrt{d_k}}\right)V.
\end{equation}
Here, $X$ is the input, $W_i^Q$, $W_i^K$, and $W_i^V$ are the weight matrices for the $i$-th head's query, key, and value projections, respectively, $W^O$ is the output projection matrix, and $d_k$ is the dimensionality of the keys. 
The FFN layer consists of several projection layers, including an up-projection, a gating mechanism, and a down-projection, which apply transformations to the input.
The major operations in both attention and FFN can be classified into two types: general matrix multiplications (GEMMs) and batched GEMMs.

During inference, LLMs generate text through two main phases: \textit{prefill} and \textit{decode}. 
The prefill phase processes the initial input sequence, while the decode phase generates output tokens auto-regressively, one token at a time, conditioning on all previously generated tokens. 
LLM decoding is memory-bound because the batch size is usually small~\cite{patel2024splitwiseefficientgenerativellm,yuan2024llm,li2024large} for GEMM operations, limiting computational resource utilization. 
Additionally, the batched GEMMs degrade to batched general matrix-vector multiplications (GEMVs) due to the auto-regressive nature of decoding. 
In this case, the matrix dimension $M$ is always 1, leading to low arithmetic intensity. 
As the context length increases during decoding, the KV cache grows linearly, imposing extensive memory capacity demands and intensifying the memory bottleneck.
Given that LLM decoding is memory-bound, compression techniques can effectively alleviate memory constraints by reducing the size of the model weights and the KV cache without substantially impacting performance. 
By applying efficient compression methods, we can decrease memory bandwidth requirements and improve inference speed.
\begin{figure*}
    \centering
    \includegraphics[width=\linewidth]{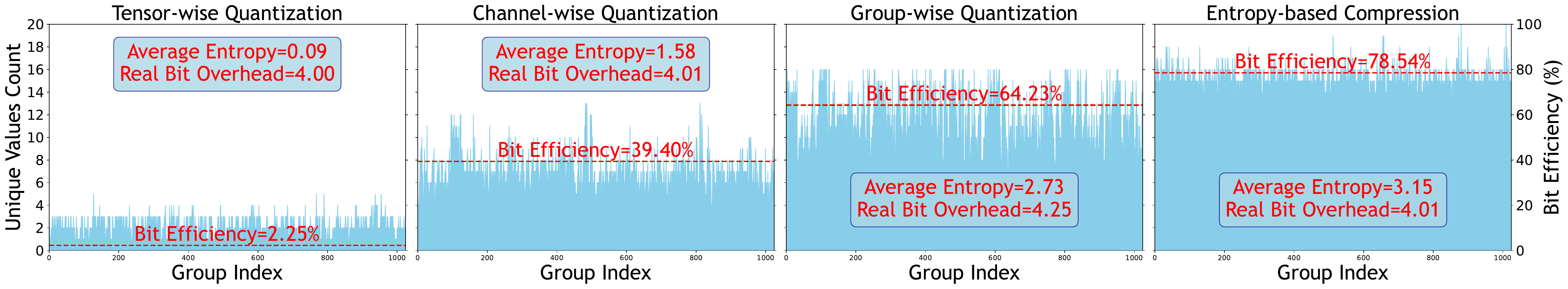}
    \caption{Unique value counts and bit efficiency across different compression methods.}
    \label{fig:entropy}
    \Description{Unique value counts and bit efficiency across different compression methods.}
\end{figure*}

\noindent \textbf{Takeaway 1:} Compression mitigates bandwidth and capacity requirements in LLM decoding.

\subsection{Compression and Information Entropy}

Quantization~\cite{han2015deep, guo2022ant, guo2022squant, lin2024awq, lin2024qserve, hu2025m} is a fundamental technique in model compression that reduces the precision of neural network weights by mapping continuous values to discrete levels. This process can be mathematically described as:
\begin{equation}
\tilde{w} = \text{Quantize}(w, S, Z) = S \cdot Q\left(\frac{w}{S} + Z\right),
\end{equation}
where $w$ represents the original weights, $\tilde{w}$ are the quantized weights, $S$ is the scaling factor, $Z$ is the zero point, and $Q(\cdot)$ denotes the quantization function.
The granularity at which quantization is applied significantly impacts both the compression ratio and the model's performance. 
Quantization can be performed at various levels, such as tensor-level, channel-level, or group-level. As illustrated in~\Cref{fig:entropy}, finer-grained quantization---from tensor to channel to group---results in a greater number of unique quantized values within each group. 
This increased diversity of quantized values implies that the quantization codes are utilized more efficiently, capturing more information and leading to less information loss.

To quantify the effectiveness of different quantization approaches, we employ the concept of information entropy $H$, defined as:
\begin{equation}
H = -\sum_{i=1}^{n} p_i \log_2 p_i,
\end{equation}
where $p_i$ is the probability of the $i$-th quantized value occurring and $n$ is the total number of unique quantized values. 
A higher entropy indicates a richer representation of information in the quantized data, which often correlates with improved model performance.
However, finer-grained quantization schemes introduce additional metadata overhead, such as an increased number of scaling factors and quantization parameters. 
This extra information can offset the benefits gained from the higher entropy of the quantized weights. 
To account for this trade-off, we define a metric called \textbf{bit efficiency} $\eta$, which normalizes the information entropy by the actual bit overhead incurred:
\begin{equation}
\eta = \frac{H}{B_{\text{real}}},
\end{equation}
where $B_{\text{real}}$ represents the real bit overhead, including both the quantized data and the associated metadata.

Improving bit efficiency can be approached in two ways: by increasing the entropy $H$ or by reducing the real bit overhead $B_{\text{real}}$. 
Compression techniques like Huffman coding aim to minimize bit overhead by assigning shorter codes to more frequent quantized values, theoretically achieving the lowest possible bit usage. 
However, in practical hardware implementations, the savings from Huffman coding may not be fully realized due to constraints like data alignment requirements. 
To mitigate this inefficiency, we can pad the bitstream with additional data, such as metadata and outliers, effectively increasing the entropy without incurring significant extra cost.
As shown in~\Cref{fig:entropy}, our proposed method achieves higher entropy and lower bit overhead compared to existing group-based quantization techniques like AWQ~\cite{lin2024awq}. 
By optimizing the balance between information content and storage efficiency, our approach enhances bit efficiency, leading to better model performance without a proportional increase in storage requirements.

% \noindent \textbf{Takeaway 2:} Enhancing information entropy through efficient compression methods directly improves bit efficiency in quantized data.

\noindent \textbf{Takeaway 2:} The primary goal of compression is to increase bit efficiency; enhancing information entropy improves model performance, while reducing real bit overhead decreases storage needs.

\subsection{Runtime De-/compression Overhead}

\begin{figure}[h]
    \centering
    \includegraphics[width=\linewidth]{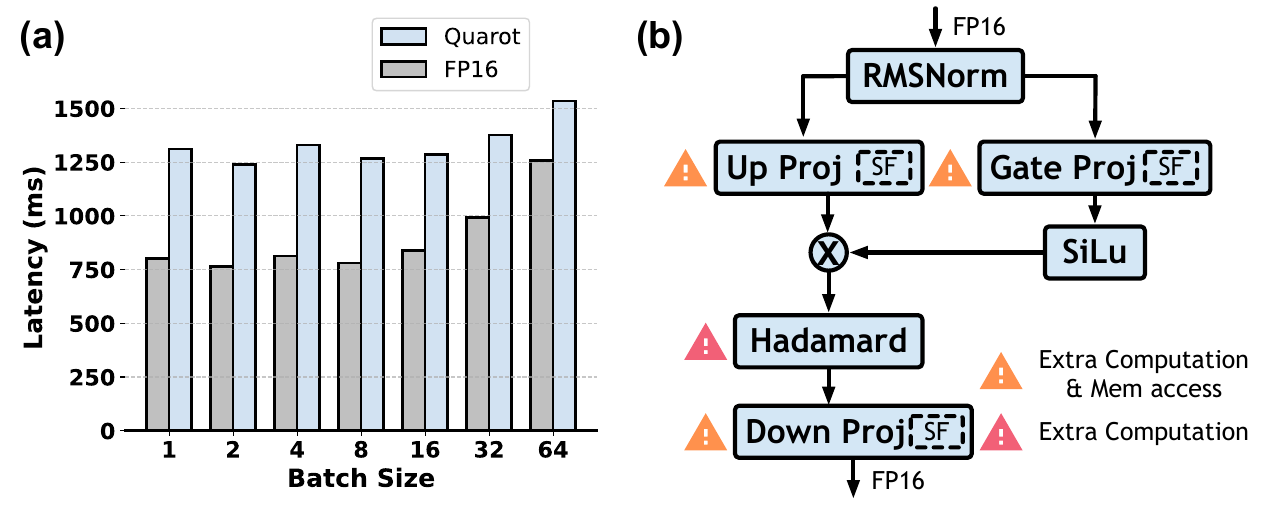}
        \caption{(a) Latency comparison between FP16 and Quarot; (b) Overhead incurred in Quarot's FFN.}
    \label{fig:runtime_overhead}
    \Description{(a) Latency comparison between FP16 and Quarot; (b) Overhead incurred in Quarot's FFN.}
\end{figure}

To achieve superior model performance in compressed neural networks, complex compression schemes are often adopted to increase information entropy. 
One such method is Quarot~\cite{quarot}, which achieves state-of-the-art accuracy in compressed LLM models by employing advanced quantization techniques. 
By rotating the model and suppressing outliers, Quarot maximizes the utilization of quantization codes, thereby enhancing information entropy while preserving model accuracy.
During runtime, as depicted in \Cref{fig:runtime_overhead}-b, Quarot's decoding process involves decompressing the weights, activations, and KV cache, as well as compressing the KV cache using compute-intensive Hadamard transformations and scaling factors. 
These steps introduce significant computational and memory overhead due to the element-wise multiplications and the need to load additional scaling factors and other metadata from memory.

As illustrated in \Cref{fig:runtime_overhead}-a, when testing on a 4-bit compressed LLaMA2-7B model with an input sequence length of 1024 and 512 decoding steps, the decoding process is approximately 0.6\x slower than the original FP16 model. 
The runtime compression and decompression overhead outweigh the saved memory bandwidth and even shift the bottleneck from memory to compute resources.
While methods like Quarot effectively reduce the storage capacity required for model weights, they do not necessarily yield inference speedups. 
The added compute and memory transaction overhead can result in slower performance, negating the benefits of reduced memory bandwidth and introducing new computational constraints.

\noindent \textbf{Takeaway 3:} Complex compression schemes can improve model accuracy but may introduce significant runtime compression and decompression overhead; 
thus, the efficiency of compressors and decompressors is critical for achieving actual speedups.

\subsection{Cache Compression}

Cache compression has been a significant area of research in computer architecture, aiming to reduce memory bandwidth and capacity requirements without incurring substantial performance penalties. 
Early representative works, such as Base-Delta-Immediate (BDI) compression~\cite{pekhimenko2012base}, achieve compression by representing data as a base value plus small deltas. 
This method effectively compresses cache lines where data values are similar, reducing redundancy and saving space.
More recent studies, like Buddy Compression~\cite{buddy}, explore innovative approaches to memory compression. 
Buddy Compression splits each compressed memory entry between high-speed GPU memory and a slower but larger disaggregated memory pool or host CPU memory. 
Highly compressible memory entries are accessed entirely from GPU memory, while incompressible entries source some of their data from off-GPU memory. 
This design balances performance and capacity by leveraging the compressibility of memory entries.

However, these general-purpose lossless compression schemes typically offer only low compression ratios and are not well-suited for LLM scenarios. 
LLMs require high compression ratios to manage the substantial memory demands of the KV cache and model weights. 
Lossless methods may not provide sufficient compression to alleviate the memory bottleneck in LLM inference.
Some studies have attempted to address activation compression using image-based techniques. 
For example, JPEG-ACT~\cite{jpeg} treats activation maps as images and employs a modified JPEG~\cite{wallace1992jpeg} algorithm to compress them. 
While effective in certain contexts, such methods are not suitable for cache line compression due to their block-based nature and computational complexity, which can introduce significant overhead and latency.

Given the specific needs of LLMs, high compression ratios and lossy compression schemes are acceptable, provided they maintain acceptable levels of model accuracy. 
Cache compression equipped with customized, high-throughput compressors and decompressors can flexibly perform compression and decompression on the fly, introducing minimal computational overhead. 
By integrating these compressors into the memory hierarchy, we can minimize extra memory transactions and reduce latency.
Furthermore, this approach can transform irregular data loads—where data and metadata such as scaling factors are separated—into regular, consecutive data block loads. Regularizing memory access patterns enhances performance by improving cache utilization and reducing access latency.

\noindent \textbf{Takeaway 4:} Specialized cache compression techniques with efficient, customized compressors and decompressors are crucial for LLMs to achieve high compression ratios with minimal overhead, effectively mitigating memory bandwidth and capacity bottlenecks.

\section{Entropy-aware Cache Compression}

\subsection{Compression Target} \label{sec:compression_target}

In our methodology for cache compression, we prioritize the preservation of memory bandwidth and efficient utilization of memory size. 
To achieve these goals, we implement a fixed compressed block size and power-of-two compression ratios. 

\noindent \textbf{Target Compressed Block Size.}
The selection of an appropriate compressed block size is critical and must balance several factors. 
We establish that the compressed block size should exceed 32 bytes, which corresponds to one sector, as this is the minimum memory transaction unit. 
Compressing to sizes smaller than this threshold could lead to significant decompression overhead or a decline in model performance due to an excess of group-level information at this scale.
Conversely, we determine that the compressed block size should not exceed 128 bytes, which is equivalent to one cache line. 
Larger granularities may reduce memory load flexibility, and expansive compressed block sizes encompass more data per block, potentially obscuring important group-level data features and consequently degrading model performance.
Considering that the default transaction size from DRAM to L2 cache is 64 bytes \cite{gtc}, we opt for a compressed block size of \textbf{64 bytes}. 
This choice is optimal for minimizing hardware overhead and maximizing bandwidth usage while storing an appropriate amount of compressed data in each block. 
This size strikes a balance between efficient compression and effective data representation, facilitating seamless integration with existing GPU cache hierarchies.

\noindent \textbf{Target Compression Ratio.}
We further refine our cache compression strategy by setting target compression ratios as \textbf{powers of two} as it facilitates easier alignment and more efficient hardware implementation. 
Typically, compression ratios of 2\x or 4\x represent an optimal balance between effective data compression and preservation of model performance.
For weights and KV cache, we target a 4\x compression ratio, resulting in an average of 4 bits per data point with 128 data points per group. 
This aggressive compression exploits the static nature of weights during inference, allowing for thorough offline analysis and optimization of the quantization process. 
For KV cache, this ratio leverages its reduced sensitivity to small errors in key and value representations. 
Conversely, for activations, we employ a more conservative 2\x compression ratio, averaging 8 bits per data point with 64 data points per group. 
This approach accounts for the dynamic range of activations during inference. 
These tailored power-of-two ratios optimize memory efficiency and model fidelity across different LLM components by considering their unique characteristics and sensitivities, significantly increasing cache capacity for weights and KV cache while maintaining computational accuracy for activations.

\begin{figure*}
    \centering
    \includegraphics[width=\linewidth]{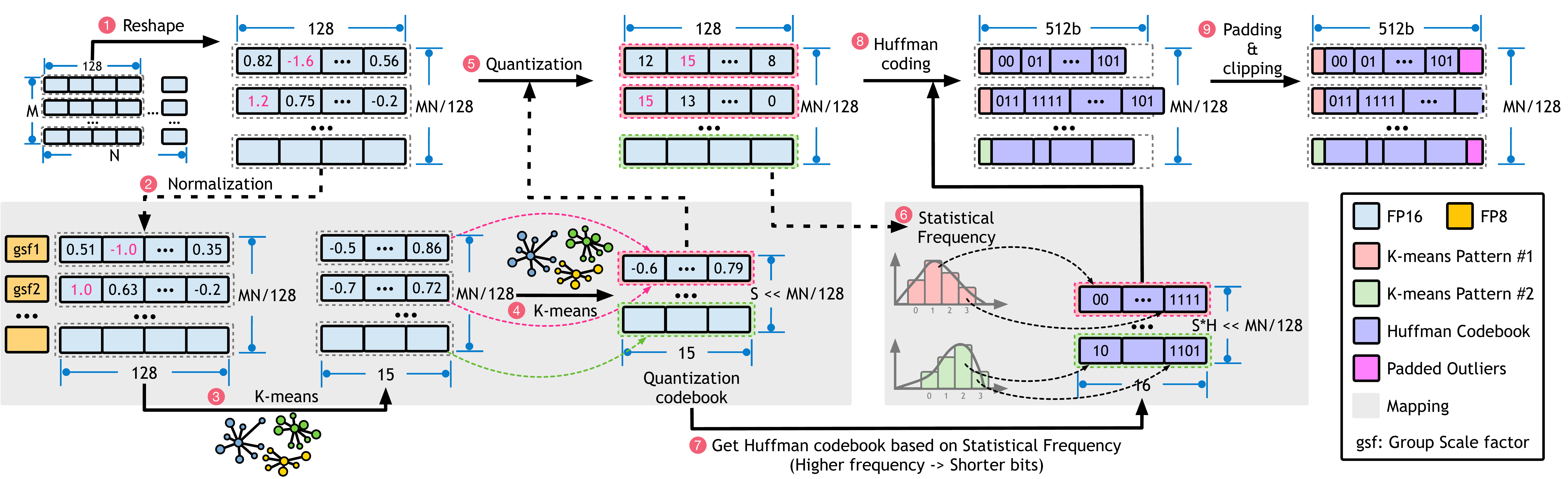}
    \caption{Overview of \methodname compression algorithm.}
    \label{fig:algo}
    \Description{Overview of \methodname compression algorithm.}
\end{figure*}

\subsection{Model Compression} \label{sec:model_compression}
\noindent \textbf{Compression Requirements.}
The cache compression target outlined in the previous section necessitates that \textbf{all group-wise information be compressed to a fixed compressed block size. }
This requirement is crucial for maintaining efficient memory transactions and avoiding additional computational overhead during decompression, which occurs on the critical data path of data loading.
To meet these requirements, our compression scheme adheres to several key principles:
\aptLtoX[graphic=no,type=html]{\begin{itemize}
    \item All group-wise data, including quantized values and associated metadata (e.g., outliers and scale factors), must be stored within the target compression granularity. 
    This approach eliminates the need for extra memory transactions during decompression.
    \item Shared metadata is utilized across multiple compressed blocks, typically at the tensor or channel level, to maximize efficiency.
    \item Decompression is designed to occur on-the-fly, with all data in the compressed block processed simultaneously. 
    This strategy precludes the need for additional compute kernels and maintains computational efficiency.
\end{itemize}}{\setlist{nolistsep}
\begin{itemize}[leftmargin=*]
    \item All group-wise data, including quantized values and associated metadata (e.g., outliers and scale factors), must be stored within the target compression granularity. 
    This approach eliminates the need for extra memory transactions during decompression.
    \item Shared metadata is utilized across multiple compressed blocks, typically at the tensor or channel level, to maximize efficiency.
    \item Decompression is designed to occur on-the-fly, with all data in the compressed block processed simultaneously. 
    This strategy precludes the need for additional compute kernels and maintains computational efficiency.
\end{itemize}}
Carefully balancing these constraints is essential to avoid complicated memory load patterns and substantial compute overhead, which would otherwise undermine the overall compression objective.

\noindent \textbf{Weight Compression.}
Existing uniform quantization approaches, such as AWQ~\cite{lin2024awq}, typically employ a group size of 128 and store an FP16 scale factor and an FP16 zero point, resulting in 32 bits of overhead. 
Storing these within the compressed block necessitates the eviction of quantized data, which is problematic for low-bit quantization scenarios. 
For instance, in 4-bit quantization AWQ, 8 original data points (32 bits overhead / 4-bit) per group of 128 would be eliminated, leading to a 6.25\% reduction in weights. 
This significant loss of information can severely impact model performance.
Non-uniform k-means quantization is generally considered to offer superior performance compared to uniform quantization. 
Current methods like SqueezeLLM \cite{kim2023squeezellm} utilize per-channel group sizes and per-tensor outliers, as the overhead of storing k-means centroids per group is substantial. 
Extracting global outliers helps preserve large values and improves k-means fitting. 
However, this approach typically requires implementation in two parts: a dense kernel for normal values and a sparse kernel for outliers. 
This dense-and-sparse approach incurs additional computational overhead, resulting in lower throughput and higher latency. 
Furthermore, storing outliers within the compressed block is challenging, as each outlier requires both location and value information. 
This leads to further data eviction, and the variable number of outliers per block can often exceed the storage capacity of a compressed block.

We propose a novel compression method that combines group-wise non-uniform quantization using predefined shared k-means patterns with clipped/padded Huffman coding. 
This approach aims to merge the strengths of integer-based uniform and k-means-based non-uniform methods while avoiding extra memory transactions or separate compute operations.
The compression process is illustrated in Figure \ref{fig:algo}.
The process begins in Step \raisebox{-0.5ex}{\includegraphics[height=1.em]{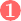}} by partitioning the tensor into $\frac{MN}{128}$ groups, each containing 128 data elements. 
In Step \raisebox{-0.5ex}{\includegraphics[height=1.em]{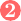}}, the data in each group undergoes a two-level normalization process. 
First, the absolute maximum value within each group is identified and normalized using a per-tensor FP16-to-FP8 scale factor, with the resulting normalized FP8 value serving as the group scale factor. 
Subsequently, the values in each group are normalized to the range $(-1, 1)$ using this FP8 scale factor. 
In Step \raisebox{-0.5ex}{\includegraphics[height=1.em]{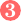}}, excluding the absolute maximum value, activation-aware k-means clustering with 15 clusters is performed on the remaining 127 values in each group. 
The sorted centroids obtained from this process are stored, forming a structure referred to as the k-means pattern. 
Moving to Step \raisebox{-0.5ex}{\includegraphics[height=1.em]{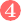}}, a second k-means clustering with $S$ clusters is applied to the k-means patterns across all groups, producing $S$ shared k-means patterns, where $S\ll\frac{MN}{128}$.
Afterwards, in Step \raisebox{-0.5ex}{\includegraphics[height=1.em]{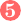}}, quantization is performed. 
The data in each group is rounded to the nearest value within each shared k-means pattern, and the pattern minimizing the output mean squared error (MSE) is selected. 
The index of the selected shared k-means pattern, along with the indices of the centroids to which the data is rounded, are recorded. 
Notably, the scale factor (i.e., the absolute maximum value in each group) is assigned index 15. 
With 15 centroids per k-means pattern, there are 15 indices for the group's data and a unique index for the scale factor, resulting in a total of 16 unique indices per group.
This method offers several advantages over previous approaches like AWQ and SqueezeLLM. 
It retains only a scale factor per group and shared k-means patterns per tensor, eliminating the need for a zero point at the group level compared to AWQ. 
It also maintains far fewer k-means patterns than SqueezeLLM while preserving superior performance.

\begin{figure}[b]
    \centering
    \includegraphics[width=\linewidth]{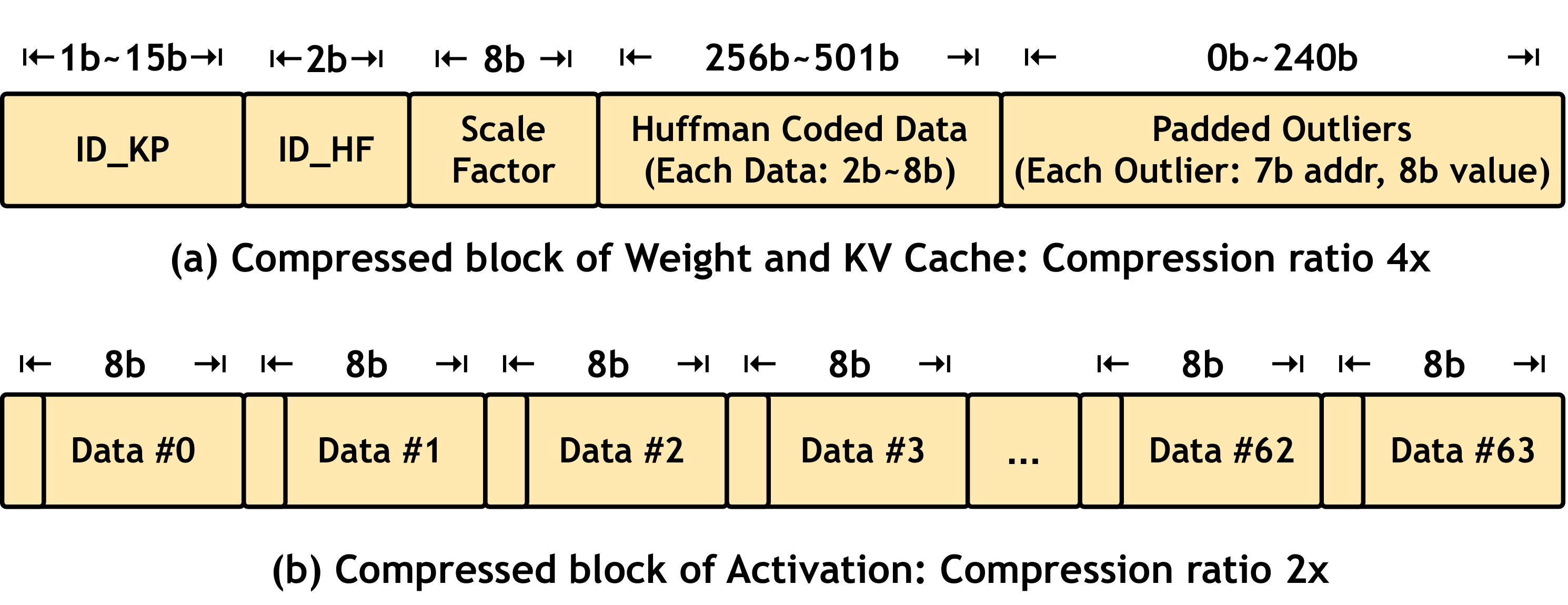}
    \caption{Compressed block format of weight, KV cache, and activations.}
    \label{fig:blocks}
    \Description{Compressed block format of weight, KV cache, and activations.}
\end{figure}

To further optimize our compression scheme, we employ clipped and padded Huffman coding to compress all information within each group. 
We observed that the index distribution within each group after quantization is highly imbalanced, with certain indices being selected significantly more frequently than others. 
Interestingly, even among groups that utilize the same shared k-means pattern for quantization, the index distribution varies considerably between groups.
To address this variability and enhance compression efficiency, we introduce $H$ Huffman codebooks, each containing 16 codes corresponding to the unique indices per group after quantization. 
This method incurs minimal overhead, requiring an additional \(log_2H\) bits to record the specific Huffman codebook used for compressing each group.
The derivation of these Huffman codebooks follows a two-step process, as illustrated in Steps \raisebox{-0.5ex}{\includegraphics[height=1.em]{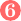}} and \raisebox{-0.5ex}{\includegraphics[height=1.em]{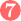}}.
First, for each shared k-means pattern, we apply k-means clustering with $H$ clusters to the frequency distributions of indices obtained after quantization. 
This clustering yields $H$ representative frequency distributions, which are then converted into $H$ Huffman codebooks. 
Finally, as depicted in Step \raisebox{-0.5ex}{\includegraphics[height=1.em]{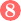}}, each quantized group selects the Huffman codebook that provides the optimal compression ratio, ensuring efficient encoding of the group data.
After compression, each group consists of four components: a fixed-length scale factor, a fixed-length Huffman codebook choice, a variable-length shared k-means pattern index, and variable-length Huffman-coded quantized data, as shown in \Cref{fig:blocks}a.
To ensure that the compressed data fits within the designated block size, we employ a clipping and padding strategy, as depicted in Step \raisebox{-0.5ex}{\includegraphics[height=1.em]{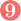}}.
If the compressed data exceeds the block size, we simply clip the excess. 
Conversely, if the data is smaller than the block size, we pad the values from the second largest absolute value (largest absolute value is stored as scale factor) to smallest until the block is filled. 
Each padded value occupies 15 bits, with 7 bits for location information and 8 bits for the FP8 value, quantized using the same per-tensor FP16-to-FP8 scale factor.

To optimize our compression methodology, we conducted a comprehensive design space exploration using the LLAMA2-7B model as our benchmark. 
This exploration aimed to determine the optimal values for two key parameters: the number of shared k-means patterns $(S)$ and the number of Huffman codebooks per k-means pattern $(H)$.

\begin{figure}
    \centering
    \includegraphics[width=\linewidth]{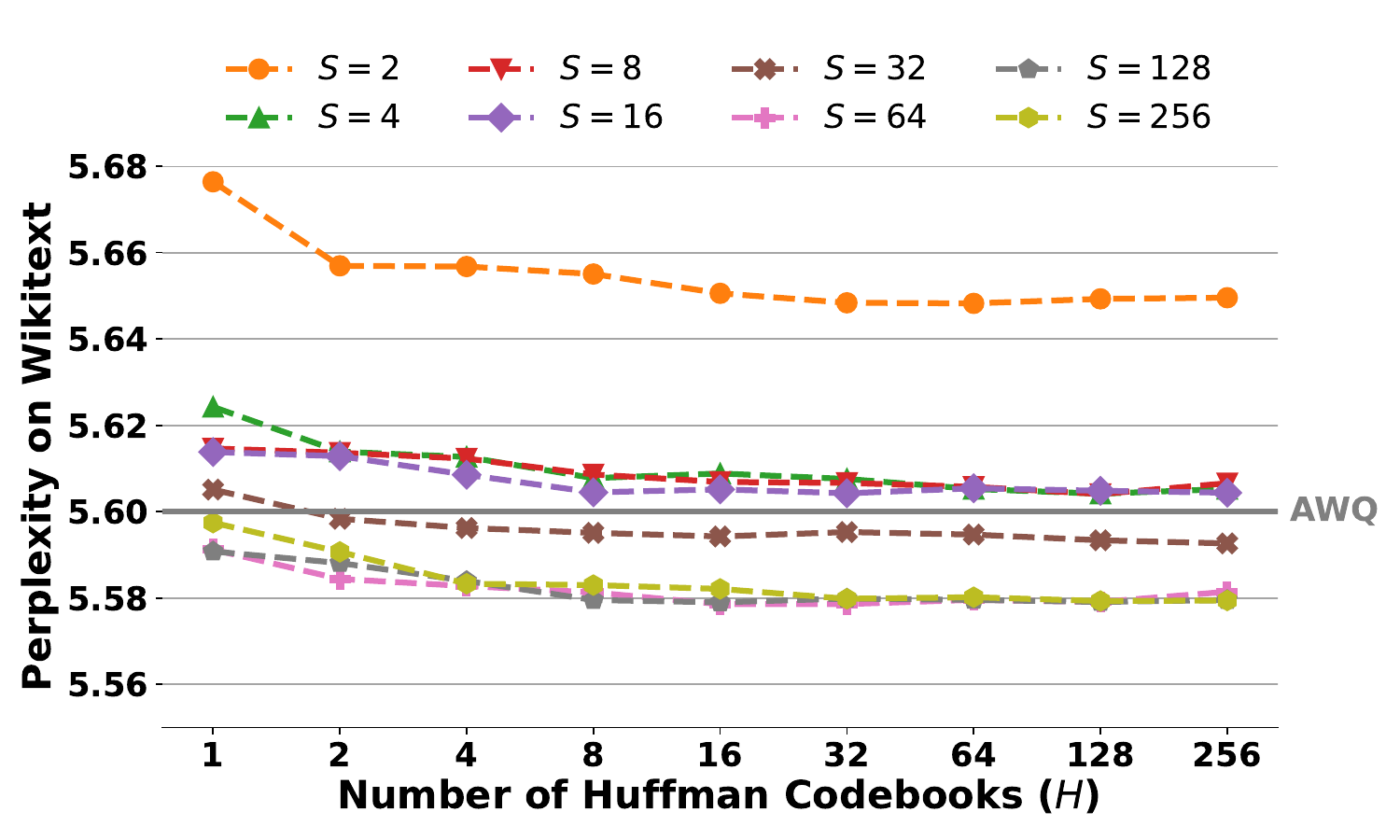}
    \caption{Design space exploration showing the impact of shared k-means patterns numbers $(S)$ and Huffman codebooks numbers$(H)$ on LLaMA2-7B perplexity.}
    \label{fig:dse}
    \Description{Design space exploration showing the impact of shared k-means patterns numbers $(S)$ and Huffman codebooks numbers$(H)$ on LLaMA2-7B perplexity.}
\end{figure}

\Cref{fig:dse} illustrates the results of our design space exploration, showing the impact of varying $S$ and $H$ on model perplexity. The exploration reveals several key insights: 
first, increasing $S$ generally improves perplexity, with diminishing returns beyond $S=64$.
Second, the impact of $H$ is less pronounced, with minimal improvements beyond $H=4$ for most $S$ values.
Based on these observations, we chose $S=64$ and $H=4$ as our optimal configuration. 
This selection balances performance improvement with implementation efficiency. 
$S=64$ provides substantial perplexity reduction compared to lower values, while avoiding the increased complexity and potential overfitting risk of larger $S$ values.
$H=4$ offers a good trade-off between compression efficiency and codebook storage overhead, as higher $H$ values show minimal additional benefit.
Furthermore, this configuration $(S=64, H=4)$ achieves perplexity better than the AWQ baseline, indicating that our method can outperform state-of-the-art performance while offering the additional benefits of our compression scheme. This choice allows us to maintain high model quality while enabling efficient hardware implementation and reduced memory footprint.

\noindent \textbf{KV Cache Compression.}
The compression of KV Cache follows a similar methodology to weight compression, with some key adaptations to accommodate its dynamic nature. 
We utilize the same calibration dataset employed in weight compression, capturing each layer's KV cache by forwarding this dataset through the model.
The offline steps (\raisebox{-0.5ex}{\includegraphics[height=1.em]{figure/redcircle1}}\raisebox{-0.5ex}{\includegraphics[height=1.em]{figure/redcircle2}}\raisebox{-0.5ex}{\includegraphics[height=1.em]{figure/redcircle3}}\raisebox{-0.5ex}{\includegraphics[height=1.em]{figure/redcircle4}}\raisebox{-0.5ex}{\includegraphics[height=1.em]{figure/redcircle6}}\raisebox{-0.5ex}{\includegraphics[height=1.em]{figure/redcircle7}}) in KV cache compression are identical to those used for weight compression. However, the final steps (\raisebox{-0.5ex}{\includegraphics[height=1.em]{figure/redcircle8}}\raisebox{-0.5ex}{\includegraphics[height=1.em]{figure/redcircle9}}) are transitioned to an online execution phase.
A notable difference arises in the selection of the shared k-means pattern (\raisebox{-0.5ex}{\includegraphics[height=1.em]{figure/redcircle5}}).
Unlike weight compression, which is static and allows for offline optimization, KV Cache compression must be performed online during runtime. 
This real-time requirement presents a unique challenge, as computing mean squared error (MSE) to find the optimal pattern for each group would incur prohibitive computational overhead in circuit implementation.
The naive approach of rounding values using all $S$ shared k-means patterns, calculating MSE for each, and selecting the minimum would be extremely costly.
For instance, processing a single group would require $G$ binary searches, $G$ subtractions, multiplications, and additions for each of the S patterns, where $G$ is the number of elements in each group.
Implementing this would either increase latency by a factor of $S$ or necessitate $S$-fold hardware duplication, both of which are impractical solutions.

\begin{figure}[h]
    \centering
    \includegraphics[width=\linewidth]{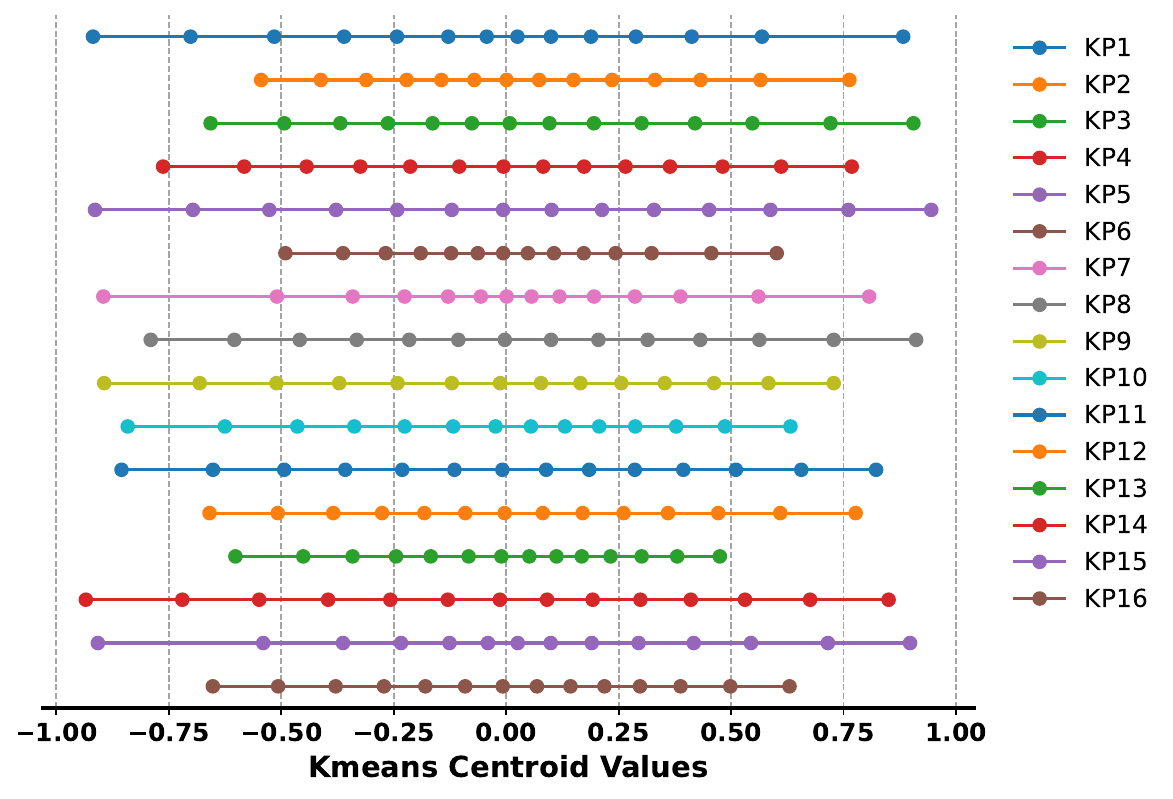}
    \caption{Highly skewed shared k-means patterns. Each row represents a k-means pattern, each point represents a centroid value.}
    \label{fig:patterns}
    \Description{Highly skewed shared k-means patterns. Each row represents a k-means pattern, each point represents a centroid value.}
\end{figure}

% how we solve the problem
Our analysis revealed that the k-means patterns are highly skewed due to the scaling of each group's k-means using the absolute maximum value (absmax), which is treated as a separate centroid and excluded from the k-means pattern. 
To visualize this phenomenon, we plotted 16 shared k-means patterns in \Cref{fig:patterns}.
This characteristic provides an opportunity for optimization.
We propose a simplified method that compares only the minimum and maximum values of each group (excluding the absmax) with the maximum and minimum of each k-means pattern. 
This approach dramatically reduces hardware complexity from $G$ comparisons to just 2, offering significantly improved flexibility in implementation.
While this approach may slightly reduce compression efficiency compared to Huffman coding, it provides more predictable compression ratios and greater adaptability to changing pattern distributions during inference.
Experimental results demonstrate that this simplified method incurs only a minimal drop in perplexity, representing an excellent trade-off between hardware realization complexity and model performance. 
This approach allows for efficient online compression of KV Cache without significant compromise to the model's accuracy or computational efficiency.
Our adapted KV Cache compression strategy achieves hardware-efficient online compression with minimal performance drop compared to MSE-based methods. 
In addition, by sharing the same structure as weight compression, it enables unified decompression implementation for both weights and KV cache.

\noindent \textbf{Activation Compression.}
Activation compression requires a different approach due to the more dynamic nature of activation ranges.
We target an average of 8 bits per activation value to accommodate this variability.
Given that activations are typically consumed by the subsequent kernel immediately after the current kernel completes, we prioritize a simple compression scheme for efficiency.
Our method employs integer-based uniform quantization with a zero point. 
The final format of a compressed block is shown in \Cref{fig:blocks}-b.
To ensure all metadata, including the scale factor and zero point, fit within the compressed block, we allocate 7 bits for the compressed data width. 
The remaining bit every 8 bits is used to store a 16-bit zero point and a 16-bit scale factor.

\section{\methodname Architecture}

\subsection{Overview} \label{sec:overview}
\methodname is designed to enhance memory system performance through advanced compression techniques, with a specific focus on GPU architectures.
\Cref{fig:system_overview} illustrates the structural components of our approach, highlighting the seamless integration of compression and decompression mechanisms within the memory hierarchy.
To optimize data transfer and storage efficiency, we have implemented a sophisticated compression system capable of operating at ratios of 2\x and 4\x.
This system comprises two key components: high-throughput low-latency parallel decompressor and compressor. 
The decompressor is strategically placed between the L1 data cache and the L2 cache, ensuring that compressed data retrieved from higher memory levels is effectively decompressed before being utilized by the streaming multiprocessors (SMs). 
Conversely, the compressor is positioned between the L2 cache and the high bandwidth memory (HBM), in close proximity to the L2 cache. 
% It compresses data as soon as it becomes available and then moves the compressed data back to the L2 Cache.
% Conversely, the compressor is positioned between the L2 Cache and the High Bandwidth Memory (HBM), compressing data as it moves from the L2 Cache to the HBM.
This arrangement follows the same high-level architecture employed in modern NVIDIA GPUs \cite{gtc, hopperinline, compressiblemem, turing}, such as Compute Data Compression \cite{cdc-a100}, making it extremely easy to integrate our compressor and decompressor into existing GPU architectures. 
% 2. why we do?
% This arrangement serves a dual purpose: it significantly reduces memory bandwidth demands and optimizes both DRAM and L2 cache capacity utilization. 
By performing compression and decompression on-the-fly, the system significantly reduces memory bandwidth demands and optimizes both DRAM and L2 cache capacity utilization.

% how to use?
% similar to NVIDIA's compressible memory \cite{compressiblemem}, 
The use of \methodname incorporates both software and hardware control. 
Users must explicitly declare compression related properties in \texttt{CUmemAllocationProp}, including whether the allocated memory is compressed and the compression ratio, prior to memory allocation.
To support this system, the page table and translation lookaside buffers (TLBs) are augmented with additional bits: one bit indicates whether a page is compressed, and another bit identifies the compression ratio.
This modification incurs no extra overhead as NVIDIA GPUs have unused bits in page table entries (PTEs). 
All explicitly declared compressible memory allocations in HBM are stored in a compressed format. 
When loading, data is fetched in its compressed form from HBM to the L2 cache; 
the L2 memory controller then directs the block to the decompressor, which decompresses the data before serving it to the SMs. 
Conversely, when writing, uncompressed data is temporarily stored in its original format.
Once all the uncompressed data required to build a compressed block is available, the L2 controller immediately directs it to the compressor, which then converts the data into compressed blocks.
By implementing this compression strategy, \methodname aims to address the critical challenges of data movement and storage during LLM inference in modern GPU architectures, leading to substantial improvements in overall system performance and energy efficiency.

\subsection{Decompressor Design} \label{sec:decompressor}

\begin{figure*}[ht]
    \centering
    \includegraphics[width=\linewidth]{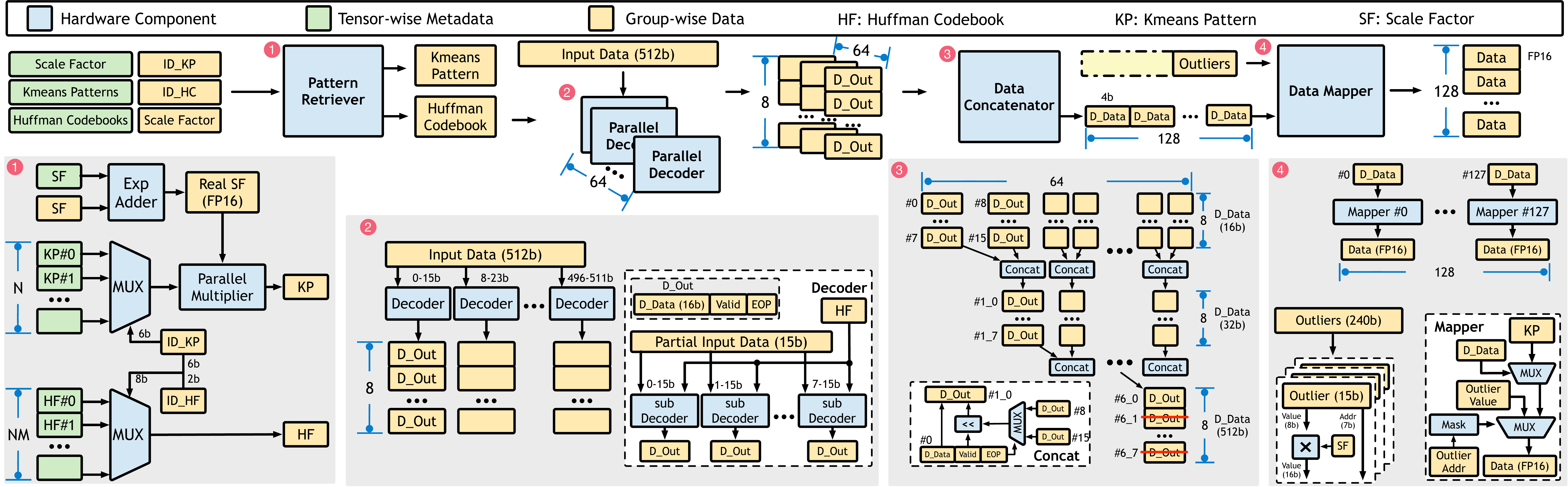}
    \caption{High compression ratio (4\x) decompressor design.}
    \label{fig:decompressor4x}
    \Description{High compression ratio (4\x) decompressor design.}
\end{figure*}

\noindent \textbf{High Compression Ratio (4\x) Decompressor.}
The decompression process for our high compression ratio (4\x) scheme involves several key steps, optimized for low-latency and high-throughput parallel processing. The details of the decompressor is depicted in \Cref{fig:decompressor4x}.
% Step 1:
Initially, prior to a compressed load operation, all tensor-wise metadata—including shared k-means patterns, Huffman codebooks, and FP16-to-FP8 scale factor is loaded into the decompression module's buffer. 
This preloaded information is only loaded once and used for the subsequent decompression of the loaded compressed block.
The decompression of a block begins with feeding the per-group scale factor, Huffman-coded shared k-means pattern choice ($ID\_KP$), and Huffman codebook choice ($ID\_HF$) into the pattern retriever, as illustrated in \raisebox{-0.5ex}{\includegraphics[height=1.em]{figure/redcircle1}}. 
% the corresponding k-means pattern and huffman codebook used be this group is retreived. and the per-group FP8 scale factor is converted to FP16 by adding the FP16-to-FP8 scale factor to the exponent, leveraging our constraint of power-of-two per-tensor scale factors.
% A Huffman decoder decodes the ID\_KP to select the appropriate k-means pattern for the group and combines decoded ID\_KP and ID\_HF to identify the corresponding huffman codebook. 
The corresponding k-means pattern and Huffman codebook used by the group are retrieved and the per-group FP8 scale factor is converted back to FP16 by adjusting the exponent, leveraging the constraint that the FP16-to-FP8 scale factor is a power-of-two.
% Concurrently, the decoder outputs the decode start position, marking the beginning of the Huffman-coded data. The per-group FP8 scale factor is converted to FP16 by adding the FP16-to-FP8 scale factor to the exponent, leveraging our constraint of power-of-two per-tensor scale factors.

% Step 2:
In the next phase, the chosen k-means pattern is multiplied by the scale factor to obtain the actual values of the k-means centroids used for group quantization. 
Simultaneously, the Huffman-coded data undergoes decoding. 
To address the inherent sequential nature of Huffman decoding while maintaining efficiency, we propose a novel parallel Huffman decoder. 
This design incorporates 64 Huffman decoders operating in parallel on overlapping sections of the sequence, as shown in \raisebox{-0.5ex}{\includegraphics[height=1.em]{figure/redcircle2}}. 
By limiting the code length to 2-8 bits, we enable efficient decoding. 
Each decoder processes 8 bits of its group ensuring decoding of at least one and at most four data points per group ($D\_Data$), plus a 7-bit overlap with the subsequent segment, ensuring the last data is correctly decoded and knowing the correct position of where the next segment should start ($EOP$).
The decoders each contain 8 sub-decoders, processing the same 15-bit data chunk but starting at different positions (0 to 7), converting Huffman codes to indices, as illustrated in \raisebox{-0.5ex}{\includegraphics[height=1.em]{figure/redcircle2}}. 
These sub-decoders operate independently in parallel until reaching the boundary of their segment's responsibility.
In this case, each decoder outputs all eight possible decoding results ($D\_Out$) corresponding to its segment, ensuring comprehensive coverage of all potential decoding paths.

% Step 3:
The result aggregation process follows, where the data concatenator in \raisebox{-0.5ex}{\includegraphics[height=1.em]{figure/redcircle3}} merges the outputs of every two neighboring decoders over six stages in a tree-like manner. 
To merge the results from two neighboring decoders, each $D\_Out$ in the left decoder's output uses its $EOP$ indicator to determine which $D\_Out$ from the right decoder should be combined with it.
Once the appropriate pairing is identified, the selected data from both decoders are concatenated and written to the next stage. 
After completing six stages of merging, only a single result remains, representing the decoding sequence starting from a specific position. 
Besides that, the concatenator also generates a mask to indicate the number of outliers padded.
% The result aggregation process follows, where the data concatenator in \redcircle{3} utilizes the decode start position from the pattern retriever to select the appropriate subsegment decoder output from the first segment decoder. 
% This process continues sequentially through the segments, shifting results into the decoded sequence until all 128 data points are processed or the last group is aggregated, depending on whether padded or clipped Huffman coding is employed.
% Step 4:
In the final step, the decoded sequence, along with the padded outliers, is mapped to the actual centroids corresponding to each index. 
The outliers are first scaled to FP16 by multiplying them with the scale factor. 
This mapping process is managed by 128 parallel mappers in the data mapper module, as depicted in \raisebox{-0.5ex}{\includegraphics[height=1.em]{figure/redcircle4}}.
Each mapper translates indices in the decoded sequence and outliers to their corresponding real centroids. 
The outlier address, combined with the outlier mask generated in the previous stage, acts as a selector. 
This selector determines whether a given position should use the value from the k-means pattern (mapped via Huffman decoding) or the scaled outlier, ensuring accurate reconstruction of the original data.
% In the final step, the decoded sequence together with outliers is mapped to the actual centroids corresponding to each index. The outlier will first be multiply with scale factor to become FP16.
% The mapping is handled by 128 parallel mappers in the data mapper module, each responsible for translating indices in the decoded sequence, and outliers to real centroids, which is shown in \redcircle{4}. The outlier address combined with the outlier mask generated from previous stage will act as selector, to select whether this should be value of outlier or the value from kmeans patterns of pos index, indicate whether each position should use the Huffman-mapped value or the padded outlier.
% The exponent adder converts padded outliers to FP16 by adding the per-tensor FP16-to-FP8 scale factor to the exponent and generates a mask to indicate whether each position should use the Huffman-mapped value or the padded outlier.
% Summary
This parallel and pipelined approach enables efficient decompression of high-ratio compressed data, balancing parallelism with the sequential nature of Huffman coding to achieve optimal performance.

\noindent \textbf{Low Compression Ratio (2\x) Decompressor.}
The 2\x decompressor employs a straightforward yet efficient approach to decompress data. 
The process begins with the extraction and combination of scale factors and zero points, which are strategically stored every 8 bits within the 512-bit compressed block. 
Concurrently, the decompressor performs a signed extension operation on the 7-bit quantized values, expanding them to 8 bits.
This extension is crucial for preserving the sign information and preparing the data for subsequent processing.
To optimize performance, the decompressor implements bit manipulation techniques proposed by those described in \cite{elephant}. 
These techniques are particularly efficient, requiring only two operations to complete the decompression process. 
The combination of localized scale factors and zero points with efficient bit manipulation allows the 2\x decompressor to achieve a balance between compression ratio and decompression speed, eliminating the need for complicated prefetching transactions.

\subsection{Compressor Design} \label{sec:compressor}

\begin{figure}[h]
    \centering
    \includegraphics[width=\linewidth]{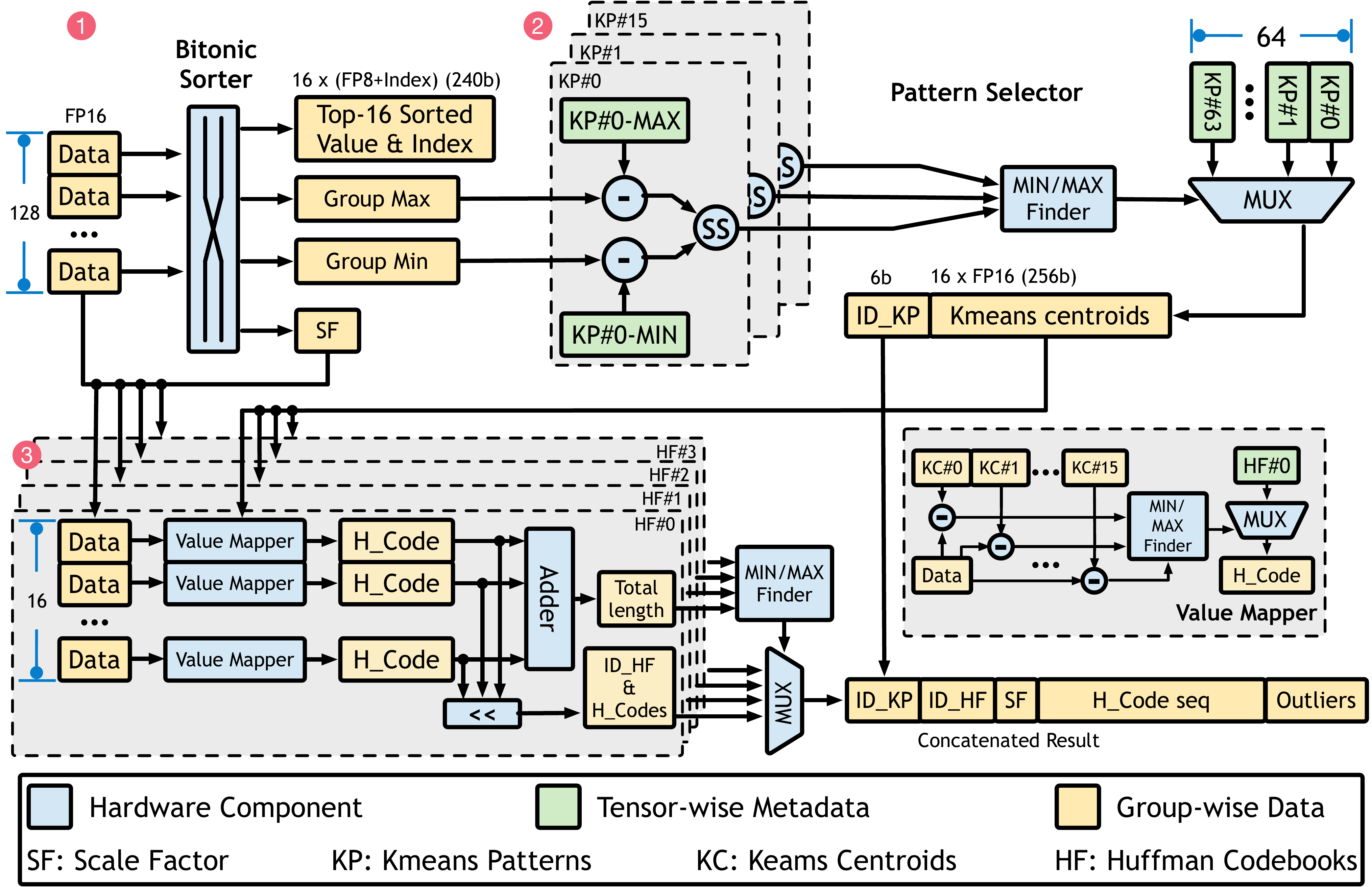}
    \caption{High compression ratio (4\x) compressor design.}
    \label{fig:compressor4x}
    \Description{High compression ratio (4\x) compressor design.}
\end{figure}

\noindent \textbf{High Compression Ratio (4\x) Compressor.}
The high compression ratio (4\x) compressor operates on 256-byte blocks, each with 128 FP16 data points. 
\Cref{fig:compressor4x} illustrates the details of the compression process. 
The process begins with a bitonic sorter, which extracts the scale factor, the top 16 sorted values and their indices for outlier padding, and the group's min/max values, as shown in \raisebox{-0.5ex}{\includegraphics[height=1.em]{figure/redcircle1}}.
Next, a pattern selector (\raisebox{-0.5ex}{\includegraphics[height=1.em]{figure/redcircle2}}) evaluates the fitness of the input data against a set of shared k-means patterns. 
To reduce overhead, the number of shared k-means patterns is decreased from 64 to 16. 
% For each pattern, the selector computes a fitness score by calculating the squared error: the difference between the pattern's minimum and maximum values and the group's minimum and maximum values is squared and summed. 
For each pattern, the selector calculates a fitness score by summing the squared differences between its min/max and the group's min/max.
This process yields $S$ fitness values, where $S$ is the number of shared patterns.
The pattern with the minimum squared error is selected for quantizing the group, and its index is recorded as $ID\_KP$.

Following pattern selection, the encoder begins parallel encoding, as depicted in \raisebox{-0.5ex}{\includegraphics[height=1.em]{figure/redcircle3}}. 
The compression process employs four encoders, each corresponding to one Huffman codebook. 
Within each encoder, parallel value mappers determine the index to which each input value should be quantized. 
This mapping is achieved by calculating the differences between the input values and the centroids of the selected k-means pattern, assigning each value to the centroid with the minimum difference.
The resulting indices are Huffman encoded and concatenated to form the total encoded sequence. 
The length of the encoded sequence is calculated, and the most efficient encoding—determined by the shortest total length—is selected. The corresponding Huffman codebook index is recorded as $ID\_HF$.
The final compressed output is constructed by concatenating the encoded sequence from the optimal Huffman encoder, the 16 outliers identified by the bitonic sorter, and the indices $ID\_KP$ and $ID\_HF$. 
If the resulting bitstream exceeds the target compressed block size, it is clipped to fit. This clipping ensures a fixed compression ratio but may result in a loss of information.

\noindent \textbf{Low Compression Ratio (2\x) Compressor.}
% the low compression ratio compressor leverages existing hardware components from the 4\x compressor, optimizing resource utilization and maintaining efficiency. 
% the low ratio compressor share the bittonic sorter to identifies the 
% to find the scale factor and zero point  of the group. and shared the same multiply and round circuit to convert fp16 to  quantized data 
The low compression ratio compressor leverages existing hardware components from the high ratio compressor, optimizing resource utilization while maintaining efficiency. 
Specifically, the low-ratio compressor shares the bitonic sorter to identify the scale factor and zero point for each group.
Additionally, it utilizes the same multiply-and-round circuit to convert FP16 data into quantized representations, streamlining the design and reducing hardware redundancy.
The final stage of compression incorporates an intricate bit-interleaving process, where the quantized values are combined with the scale factor and zero point. 
This interleaving is specifically designed to facilitate fast dequantization during decoding, enabling efficient reconstruction of the original data while minimizing computational overhead.
% The low compression ratio (2\x) compressor leverages existing hardware components from the 4\x compressor, optimizing resource utilization and maintaining efficiency. 
% Initially, the min/max finder identifies the maximum and minimum values within the input group, establishing the data range. 
% The compressor then repurposes the sum of squares unit from the pattern selector in \Cref{fig:compressor4x}-b to calculate the zero point. 
% Quantization is achieved by reusing the parallel sum of square unit from the 4\x compressor's pattern selector. 
% These units are employed in four successive passes to generate the quantized results, effectively balancing hardware reuse with compression requirements. 
% The final stage of compression involves an intricate bit-interleaving process, where the quantized values are combined with the scale factor and zero point. 
% This is accomplished by distributing individual bits of the scale factor and zero point at 8-bit intervals throughout the compressed data stream, ensuring that local scaling information is readily available during decompression. 
% This approach not only achieves the target 2\x compression ratio but also maintains low-latency access to critical compression parameters, facilitating efficient decompression. By reusing hardware components and implementing this bit distribution strategy, the 2x compressor achieves a balance between compression efficiency and hardware efficiency.
\section{Evaluation}

\begin{table*}[!ht]
\centering
\caption{Perplexity Comparison of Models Under Different Configurations on WikiText-2 with 2048 Sequence Length.}
% \small
\begin{tabular}{llccccccc}
\toprule
  \multicolumn{2}{l}{Perplexity ↓} 
& \multicolumn{3}{c}{LLaMA} 
& \multicolumn{3}{c}{LLaMA-2} 
& Mistral \\
\cmidrule{1-9}
Bits         & Method  
&   LLaMA-7B    &   LLaMA-13B   &   LLaMA-30B   &   LLaMA-2-7B    &   LLaMA-2-13B   &   LLaMA-2-70B   &   Mistral-7B \\
\midrule
{\makecell[l]{FP16}} & -      
&  5.68   &  5.09   &  4.10   &  5.47   &  4.88   &  3.32   &  5.25 \\
\midrule
\multirow{4}{*}{\makecell[l]{W4A16 \\ g128}} 
& GPTQ-R  
&  5.83   &  5.20   &  4.22   &  5.63   &  4.99   &  3.43   &  5.39 \\
& Olive
&  6.04   &  5.38   &  4.32   &  5.81   &  5.10   &  3.43   &  5.51 \\
& AWQ     
&  \textbf{5.78}   &  5.19   &  4.21   &  5.60   &  4.97   &  3.41   &  5.37 \\
& \methodname    
&  5.80   &  \textbf{5.17}   &  \textbf{4.20}   &  \textbf{5.58}   &  \textbf{4.97}   &  \textbf{3.40}   &  \textbf{5.36} \\ 
                    
\midrule
\multirow{6}{*}{\makecell[l]{W4A8KV4 \\ g128}} 
& RTN     
&  6.23   &  5.46   &  4.56   &  5.99   &  5.19   &  3.70   &  5.59 \\
& AWQ     
&  5.93   &  5.36   &  4.39   &  5.83   &  5.12   &  3.51   &  5.50 \\
& QuaRot  
&  5.91   &  5.26   &  4.30   &  5.71   &  5.06   &  3.45   &  \textbf{5.39} \\
& QoQ
&  5.89   &  5.25   &  4.28   &  5.70   &  5.08   &  3.47   &  5.42 \\ 
& \methodname    
&  \textbf{5.87}   &  \textbf{5.22}   &  \textbf{4.24}   &  \textbf{5.65}   &  \textbf{5.03}   &  \textbf{3.44}   &  5.41 \\ 
\bottomrule

\end{tabular}
\label{tab:evaluation:wikitext}
\end{table*}
% \subsection{Experiment Setup}

\subsection{Accuracy Evaluation}
\noindent \textbf{Models and Datasets. } Our comprehensive evaluation encompasses a wide range of large language models, including the LLaMA family and Mistral. 
Specifically, we assess LLaMA (7B, 13B, 30B)~\cite{touvron2023llama}, LLaMA-2 (7B, 13B, 70B)~\cite{llama2}, and Mistral 7B~\cite{mistral}. 
This diverse selection of models, ranging from 7 billion to 70 billion parameters, allows for a thorough examination of our novel compression algorithm across various model architectures and scales. 
We used a small calibration set from the Pile \cite{pileval} dataset in order not to overfit to a specific downstream domain.
We evaluate the compressed models' perplexity on WikiText-2~\cite{wiki2} and measure their zero-shot accuracy on the commonsense tasks PIQA (PQ) \cite{piqa}, ARC \cite{arc}, HellaSwag (HS) \cite{hellaswag}, and WinoGrande (WG) \cite{winogrande} using lm\_eval \cite{eval-harness}.

\noindent \textbf{Baseline Methods. }
We compare our approach \methodname against several state-of-the-art quantization baselines and configurations. Our evaluation framework is implemented using HuggingFace and PyTorch libraries and run on NVIDIA A100 GPUs.
The evaluation includes full precision FP16 as a baseline, along with two main quantization configurations: W4A16  (4-bit weights, 16-bit activations and 16-bit KV cache) and W4A8KV4 (4-bit weights, 8-bit activations and 4-bit KV cache), both with a group size of 128. 
For the W4A16 configuration, we evaluate GPTQ-R~\cite{gptq}, Olive~\cite{olive}, and AWQ~\cite{lin2024awq} methods. 
The W4A8KV4 configuration encompasses RTN, AWQ~\cite{lin2024awq}, QuaRot~\cite{quarot}, 
and QoQ~\cite{lin2024qserve} techniques. 
Note that although Olive is considered as W4A16, during its searching phase, most layers result in using 8-bit precision instead of 4-bit.
All quantization techniques, with the exception of FP16, Olive and \methodname, are sourced directly from the QServe~\cite{lin2024qserve} paper.
These baselines represent a range of quantization approaches, from full precision to various 4-bit quantization schemes, ensuring fair and compreshensive comparisons.

\noindent \textbf{Perplexity Analysis.}
The results presented in \Cref{tab:evaluation:wikitext} provide a comprehensive overview of the WikiText-2 perplexity scores for various language models and quantization methods, with a sequence length of 2048. 
Lower perplexity scores indicate better language model performance.
The compressed models, particularly our proposed \methodname method, demonstrate competitive performance while significantly reducing model size.
In the W4A16 configuration, our \methodname method demonstrates competitive performance. 
\methodname outperforms all the baselines across all benchmarks except for LLaMA-7B, achieving an average perplexity reduction of only 0.10 compared to the FP16 baseline.
The W4A8KV4 configuration presents a more challenging scenario due to its higher compression rate. 
In this configuration, \methodname continues to deliver strong performance. 
It outperforms all baseline methods, including Quarot, which represents the most advanced and complex compression algorithm, on all LLaMA and LLaMA2 models ranging from 7B to 70B. 
Additionally, \methodname surpasses all baseline methods and achieves comparable perplexity loss to Quarot on Mistral models.
This demonstrates that \methodname is particularly effective at preserving model quality even under aggressive quantization.

\noindent \textbf{Zero-shot Accuracy Analysis.}
In Table \ref{tab:evaluation:cs}, we present the zero-shot accuracy for five commonsense tasks. Across these benchmarks, \methodname outperforms other 4-bit quantization methods in every case except for ARC-c. Notably, on the Winogrande task, \methodname surpasses all baselines by more than 2\%, underscoring its superior performance.

\begin{table}[!ht]
\centering
\caption{Zero-shot accuracy on five common sense tasks.}
\small
\begin{tabular}{lcccccc}
\toprule
  \multicolumn{2}{l}{LLaMA-2-13B} 
& \multicolumn{5}{c}{Zero-shot Accuracy ↑}  \\
\cmidrule{2-7}
Method  
&   PQ    &   ARC-e   &   ARC-c   &   HS    &   WG   &   Avg.  \\
\midrule
Origin (FP16)      
& 80.52   & 77.44   & 49.06   & 79.38   & 72.22   & 71.72   \\
Quarot(W4A4)
& 78.89   & 72.98   & 46.59   & 76.37   & 70.24   &  69.01 \\
Atom (W4A4)
& 76.50   &  57.49  & 42.32   & 73.84   & 67.40   &  63.51 \\
QoQ (W4A8KV4)    
& 79.43   &  \textbf{77.06}  & 48.81   & 78.35   & 70.48   & 70.83 \\
\methodname (W4A8KV4)
& \textbf{79.82} & 77.02  & \textbf{49.49}  & \textbf{78.61} & \textbf{72.53} & \textbf{71.49}\\

\bottomrule

\end{tabular}
\label{tab:evaluation:cs}
\end{table}

% These results demonstrate that \methodname provides an effective approach to model quantization, successfully preserving model quality across diverse model sizes and architectures while achieving significant compression. 
% The method's consistent performance across all models, highlights its potential for enabling the deployment of state-of-the-art language models in resource-constrained environments without significant loss in performance.

\begin{figure}[t]
    \centering
    \includegraphics[width=\linewidth]{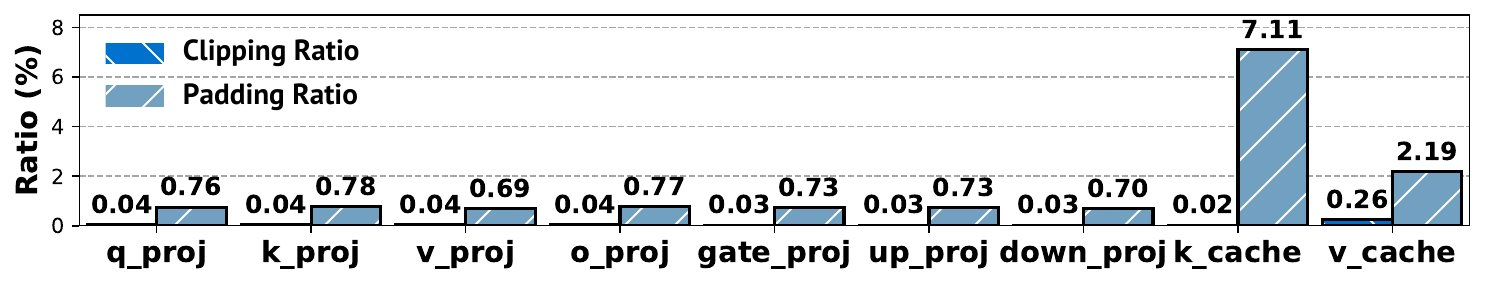}
    \caption{Average padding and clipping ratios by layer.}
    \label{fig:ratios}
    \Description{Average padding and clipping ratios by layer.}
\end{figure}

\noindent \textbf{Padding/Clipping Ratio Analysis.}
Figure \ref{fig:ratios} shows the average ratio of values clipped/padded in each layer on LLaMA2-13B. For projection layers, the clipping ratio is below 0.04\%, meaning almost no values are clipped, while the padding ratio is around 0.7\%, indicating that most outliers are preserved. It is worth noting that we achieve padding ratios of 7.11\% and 2.19\% on the K-cache and V-cache, respectively. This result demonstrates the effectiveness of our Huffman coding based compression methodology, which minimizes clipping and maximizes padding given limited space to ensure the accuracy of the compressed model.

\subsection{Area and Power Evaluation}
\noindent \textbf{Implementation.}
The fundamental RTL logic of the \methodname architecture, encompassing the compressor and decompressor, is implemented in Verilog. 
Additionally, we utilize the Synopsys Design Compiler to synthesize this RTL logic, enabling the generation of area and power statistics using ARM’s standard cell library under a commercial 28nm process. 
For a fair overhead comparison, we scale the area and power metrics to 7nm, aligning them with the specifications of the NVIDIA A100 GPU chip. 

\noindent \textbf{Result Analysis.}
We assess the power consumption and area overhead of our hardware design.
The critical path of the decompressor includes retrieving the Huffman codebook, a parallel Huffman decoder, result merging, and mapping to real values. 
By carefully designing the circuitry and inserting buffers, we achieved a pipelined design with a latency of 28 clock cycles. 
The compression latency is 62 cycles as it is not on the critical path and we trade latency for area efficiency.
To ensure that the compressor and decompressor units do not become performance bottlenecks, we replicated them 20 times, aligning their throughput with the L2 cache's peak throughput of 5120 bytes per clock cycle.
This strategy ensures seamless data compression and decompression without impeding cache performance.
Table \ref{tab:area_power} shows the area and power consumption of all duplicated instances.
Our design occupies a total area of 5.11~mm\(^2\) on the NVIDIA A100 GPU, which has a die area of 826~mm\(^2\). 
This means our hardware addition accounts for less than 1\% of the chip area, indicating a minimal area overhead. 
In terms of power consumption, our design draws a total of 7.36~W, which is less than 10\% of the A100's idle power of 82~W at 1410~MHz.
These results demonstrate that our hardware design is both efficient and practical. 
The negligible area and power overhead make it feasible to integrate into existing GPU architectures without significant alterations.

\begin{table}[b]
\centering
\caption{The area and power of \methodname on A100.}
\small
\begin{tabular}{@{}lccc@{}}
\toprule
\textbf{Component} & \textbf{Area (mm\(^2\))} & \textbf{Area Ratio} & \textbf{Power (W)} \\ \midrule
Decompressor 4\x & 3.19 & 0.39\% & 4.82 \\
Decompressor 2\x & 0.57 & 0.07\% & 0.83 \\ 
Compressor 4\x & 0.91 & 0.11\% & 1.15 \\ 
Compressor 2\x & 0.44 & 0.05\% & 0.56 \\ 
\bottomrule
\end{tabular}
\label{tab:area_power}
\end{table}

\subsection{Performance Evaluation}
\noindent \textbf{Simulator Setup.}
To evaluate the performance of our novel cache compression technique, we modified and extended Accel-Sim \cite{accelsim} and GPGPU-Sim 4.0 \cite{gpgpusim}. 
Traces were extracted using NVBit version 1.7.1 \cite{nvbit} to ensure accurate representation of program execution. 
All programs were compiled using CUDA 11.8. 
We obtained the NVIDIA A100 GPU \cite{a100} configuration by utilizing the tuner and correlated the simulated cycles with real hardware performance on a set of benchmarks provided by Accel-Sim, as well as our own benchmarks on cuBLAS \cite{cublas} and CUTLASS \cite{cutlass}. 
The simulation error is within 10\% of the real GPU performance.

\noindent \textbf{Models and Baselines.}
As LLaMA models share the same backbone within the same size across generations, we select one model if the sizes of LLaMA1 and LLaMA2 are equal. 
Our simulations include LLaMA-7B, LLaMA-13B, LLaMA-30B, LLaMA-65B \cite{touvron2023llama}, LLaMA2-70B \cite{llama2}, and Mistral-7B \cite{mistral}. We compare the performance against state-of-the-art frameworks, including TensorRT-LLM in FP16 precision \cite{trtllm}, AWQ in W4A16 \cite{lin2024awq}, SmoothQuant in W8A8 \cite{smoothquant}, and Olive \cite{olive}, where we unified all layer weights to 8-bit and adopted W8A8 precision. 
Quarot \cite{quarot} is excluded from the comparison as its performance consistently falls short of the FP16 baseline when the batch size is $\leq$ 64, as shown in Figure \ref{fig:runtime_overhead}.
We omit the prefill phase since it is compute-bound, runs only once, and occupies a negligible portion of execution time. 
Instead, our focus is on the decoding phase. 
We thoroughly evaluated our design across various batch sizes and sequence lengths to ensure comprehensive performance analysis.

\begin{figure}[!t] % or [h], [b], etc. as desired
    \centering
    
    %------------ Sub-figure (a) ------------
    \begin{subfigure}[b]{\columnwidth}
        \centering
        \includegraphics[width=\columnwidth]{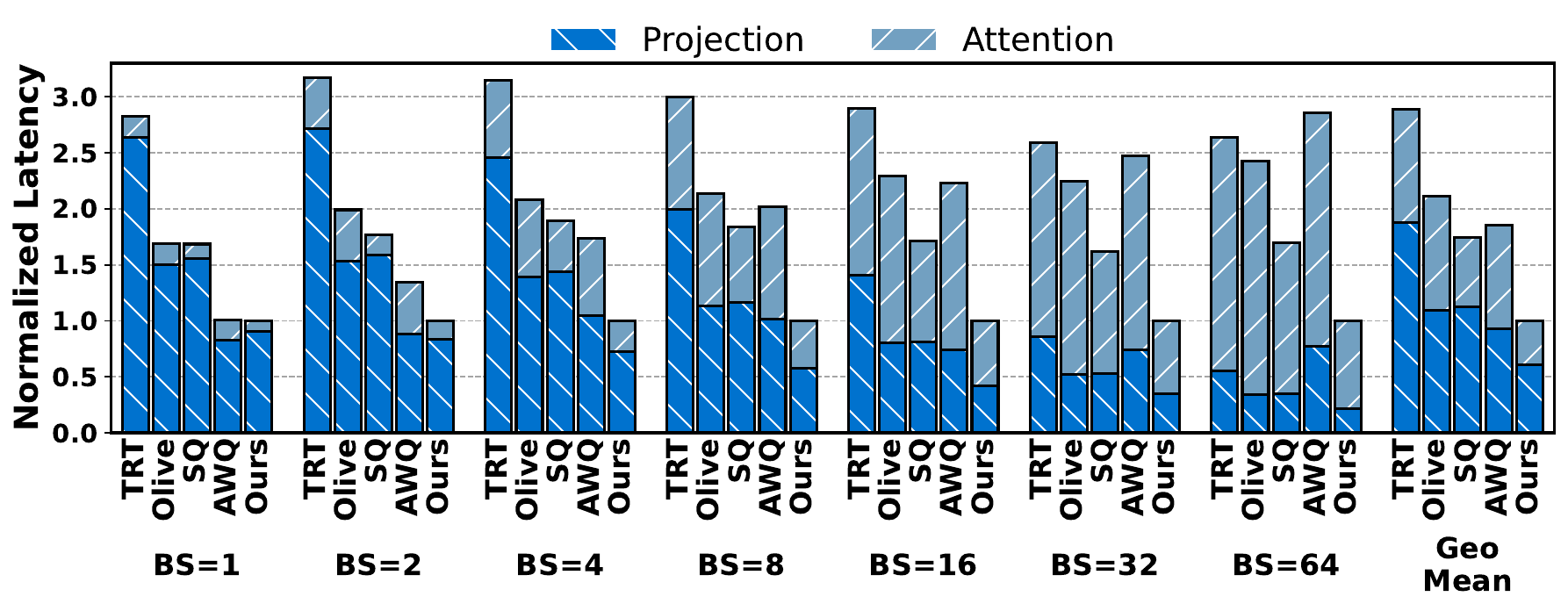}
        \vspace{-1.8em}
        \caption{\small{Normalized latency vs. batch sizes on LLaMA-13B.}}
        \label{fig:result_ba}
    \end{subfigure}
    
    % \vskip 0.5em  % space between sub-figures
    \vskip 1.5em  % space between sub-figures
    
    %------------ Sub-figure (b) ------------
    \begin{subfigure}[b]{\columnwidth}
        \centering
        \includegraphics[width=\columnwidth]{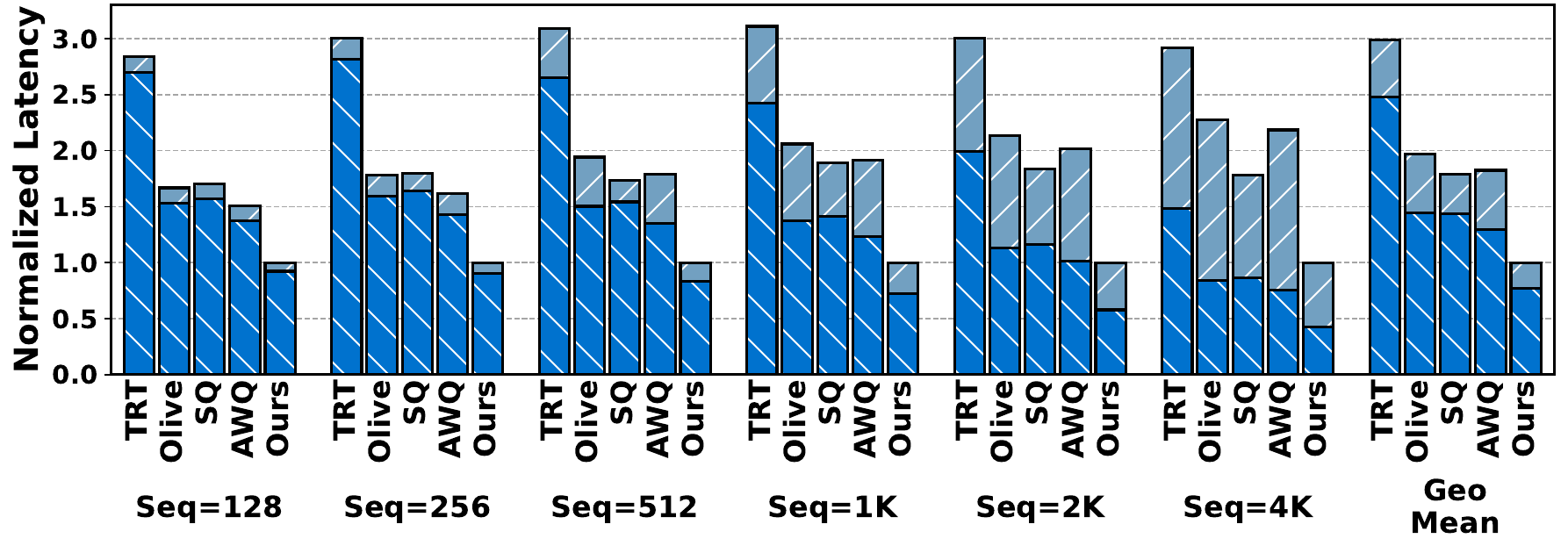}
        \vspace{-1.8em}
        \caption{\small{Normalized latency vs. sequence lengths on LLaMA-13B.}}
        \label{fig:result_seq}
    \end{subfigure}
    
    \vskip 1.5em  % space between sub-figures
    
    %------------ Sub-figure (c) ------------
    \begin{subfigure}[b]{\columnwidth}
        \centering
        \includegraphics[width=\columnwidth]{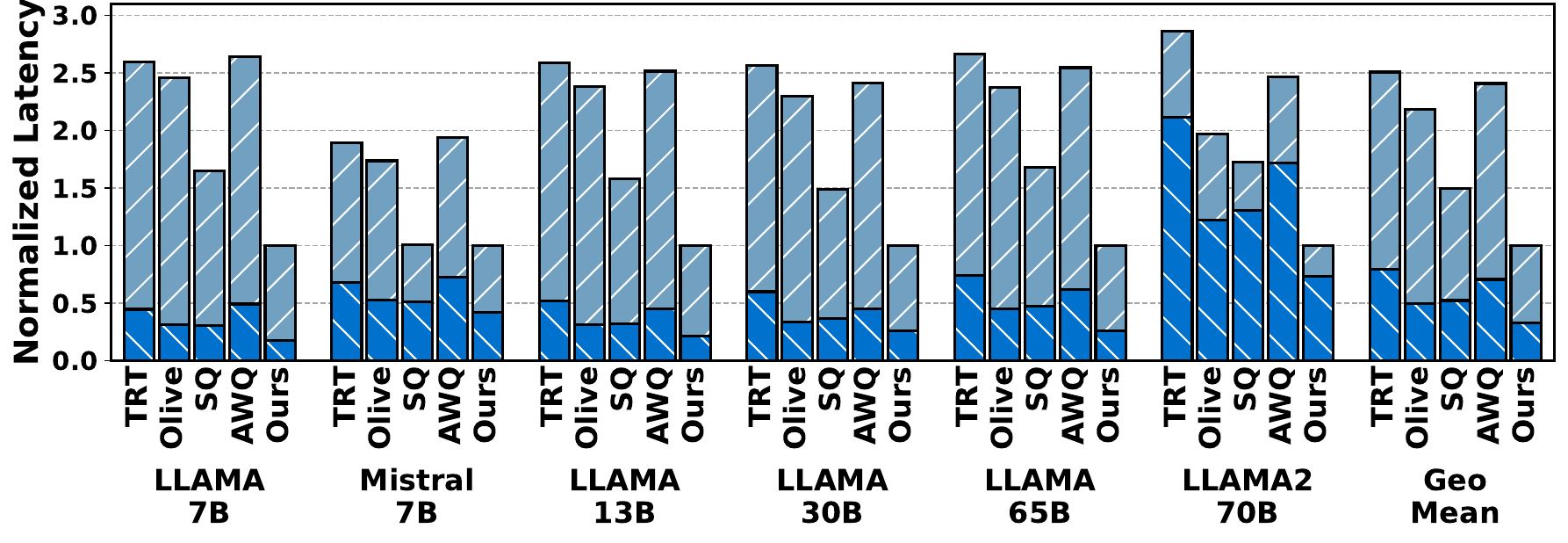}
        \vspace{-1.8em}
        \caption{\small{Normalized latency vs. various models.}}
        \label{fig:result_model}
    \end{subfigure}

    \vskip 1.0em  % space between sub-figures

    \caption{
      Comparison of normalized latency. 
      (a)~Batch size variation; 
      (b)~Sequence length variation;
      (c)~Different models.
    }
    \label{fig:combined_vertical_single_col}
    \Description{
      Comparison of normalized latency. 
      (a)~Batch size variation; 
      (b)~Sequence length variation;
      (c)~Different models.
    }
\end{figure}

\noindent \textbf{Speedup Analysis.}
We first evaluate our method on the LLaMA-13B model with sequence length 2048, sweeping batch sizes from 1 to 64.
Figure~\ref{fig:result_ba} shows that our method consistently outperforms TensorRT, Olive, SmoothQuant, and AWQ across all batch sizes. 
For projection layers, although latency increases with batch size for all methods, AWQ incurs the highest overhead while ours remains the lowest. 
Compared to FP16, our method achieves a speedup from 2.8$\times$ to 3.2$\times$ when batch size grows from 1 to 2, thanks to a reduced kernel launch latency percentage. 
Beyond this, speedup gains diminish due to the attention mechanism’s lesser improvement compared to GEMM (which is highly optimized in CUTLASS) and a reduced impact of memory bandwidth limits at larger batch sizes. 
Overall, our method attains a speedup of 2.6$\times$ to 3.2$\times$, averaging 2.9$\times$ versus TensorRT FP16, and when compared with AWQ, Olive, and SmoothQuant, provides up to 2.9$\times$, 2.4$\times$, and 1.8$\times$ speedups as sequence length increases. These results demonstrate our approach’s effectiveness in reducing latency and computational overhead, especially at larger batch sizes and sequence lengths.

We then evaluate our method across various sequence lengths with a batch size of 8 on the LLAMA-13B model.
The normalized latency compared to various baselines is shown in Figure~\ref{fig:result_seq}. 
In this scenario, the effect of kernel launch latency at small sequence lengths becomes more pronounced. 
As the sequence length increases, we observe that compared to the FP16 baseline, the speedup achieved by our method increases from 2.8$\times$ to 3.1$\times$, before starting to decrease. 
This trend is due to the diminishing impact of kernel launch overhead at larger sequence lengths, balanced against other system limitations.
When compared to AWQ, Olive, and SmoothQuant, our method consistently demonstrates increasing speedup, reaching up to 2.1$\times$, 2.3$\times$, and 1.9$\times$, respectively. 
The performance gains are more significant at longer sequence lengths, where the benefits of our optimized compression and reduced computational overhead become more evident.
We did not test sequence lengths longer than 4K due to slow simulation speeds. 
However, we expect this trend of increasing speedup to continue as the sequence length grows, given that longer sequences further amplify the advantages of our method in reducing memory bandwidth and computational demands.

% evaluate across models with batch size 32 and sequence length  4096, which is a common setting in modern inference systems. shown in figure 11. 
% out method achieves more than 2x speedup on most models.
% the mistral 7b and llama2 70b speedup is less than other models is is uses group query attenion, and the batched gemv becomes batch gemm as multiple query heads share the same one kv head. the potential allivate the memory bottlenect. 
% overall our method contantly outperforms all the baselines and achieve an average of 2.5x, 2.2x and 2.1x compared to fp16, olive and awq. 

We further evaluate our method across various models using a fixed batch size of 32 and a sequence length of 4096 tokens, settings that are common in modern inference systems. 
The models tested include LLAMA variants and other large-scale language models. 
The normalized latency results compared to the baselines are presented in Figure~\ref{fig:result_model}.
Our method achieves more than 2$\times$ speedup on most of the evaluated models. 
Notably, for Mistral-7B and LLAMA2-70B, the speedup is slightly less than that observed in other models. 
This discrepancy is attributed to the use of grouped query attention (GQA) in these models. 
GQA allows multiple query heads to share a single KV head, effectively transforming the batched GEMVs into batched GEMMs. 
This change increases the arithmetic intensity of the operations and potentially alleviates the memory bottleneck that is prominent in standard attention mechanisms.
Despite this, our method still provides significant performance gains over the baselines for all models tested. 
On average, our method achieves speedups of 2.5$\times$, 2.2$\times$, 1.5$\times$, and 2.1$\times$ compared to the FP16 baseline, Olive, SmoothQuant, and AWQ, respectively. 
These consistent improvements underscore the effectiveness of our approach in optimizing inference performance across a diverse set of models.

\begin{figure}[bt]
    \centering
    \includegraphics[width=\linewidth]{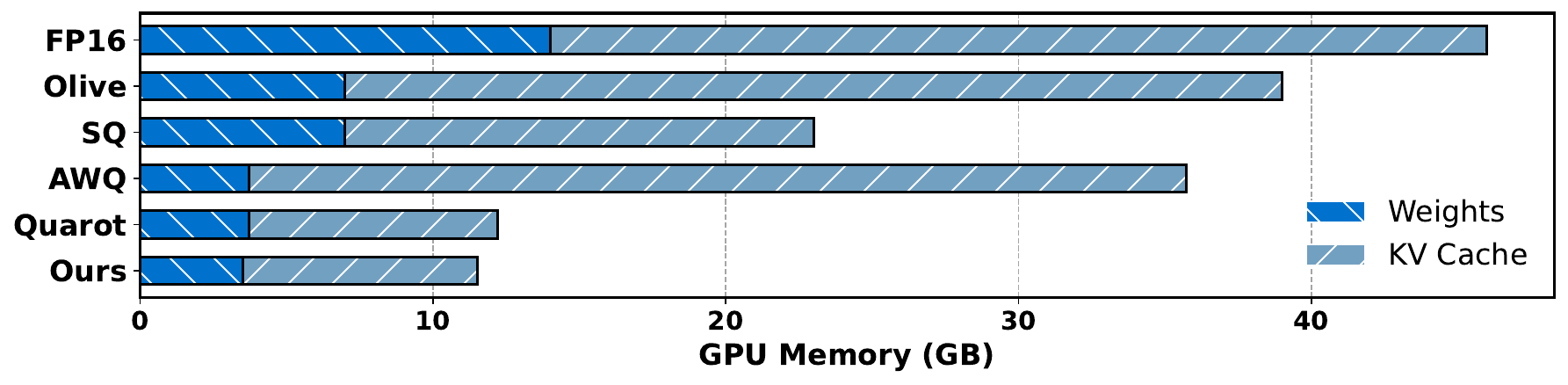}
    \vspace{-2em}
    \caption{GPU memory consumption on LLaMA-7B.}
    \vspace{-1em}
    \label{fig:result_mem}
    \Description{GPU memory consumption on LLaMA-7B.}
\end{figure}

\noindent \textbf{Memory Analysis.} Since HBM memory on GPUs is critical for running large models, we examine GPU memory consumption across various baselines on LLaMA-7B with a batch size of 32 and a sequence length of 2K. 
As shown in Figure~\ref{fig:result_mem}, our method reduces memory consumption by 3.98$\times$ compared to the FP16 baseline, which is very close to 4$\times$. 
The only overhead is a small codebook shared across tensors. 
Compared to the aggressive compression methods SmoothQuant and Quarot, our method achieves 1.99$\times$ and 1.06$\times$ reductions in memory usage, respectively.
Moreover, because our method operates on each tensor independently, it scales cleanly with no overhead across multiple GPUs. 
Therefore, when launching large models on multiple GPUs, our method can save up to 75\% of overall GPU usage.
This also translates to up to 12.8$\times$ energy savings, as we not only reduce the number of GPUs needed but also achieve up to a 3.2$\times$ speedup with a 1\% power overhead.

To further analyze the source of the observed speedup, we provide detailed memory request metrics for a GEMM kernel used in LLaMA-13B. 
Since the decoding process is highly memory-bound, the number of memory requests serves as a proxy for overall performance. 
Compared to the FP16 baseline, our method reduces memory traffic by 3.56$\times$, and the number of memory requests decreases by 1.98$\times$ and 1.28$\times$ relative to SmoothQuant and AWQ, respectively. These results demonstrate the superior performance of our approach compared to fused dequantization and matmul kernels.

\begin{figure}[bt]
    \centering
    \includegraphics[width=\linewidth]{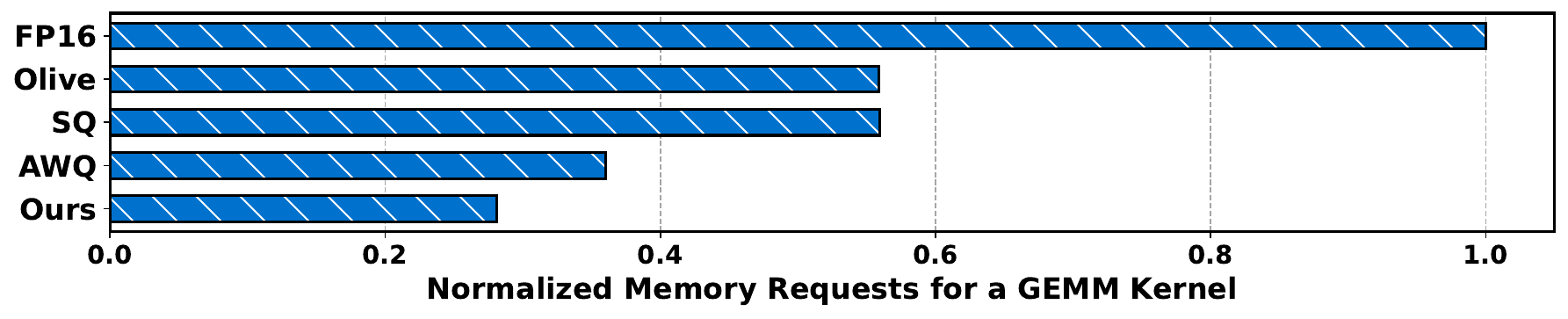}
    \vspace{-1.5em}
    \caption{Normalized Memory Requests for a GEMM Kernel (M=16, K=5120, N=13824) in LLaMA-13B.}
    % \vspace{-1em}
    \label{fig:result_request}
    \Description{Normalized Memory Requests for a GEMM Kernel (M=16, K=5120, N=13824) in LLaMA-13B.}
    \vspace{-5pt}
    
\end{figure}

\begin{figure}[!t]
    \centering
    \begin{subfigure}[b]{0.48\linewidth}
        \centering
        \includegraphics[width=\linewidth]{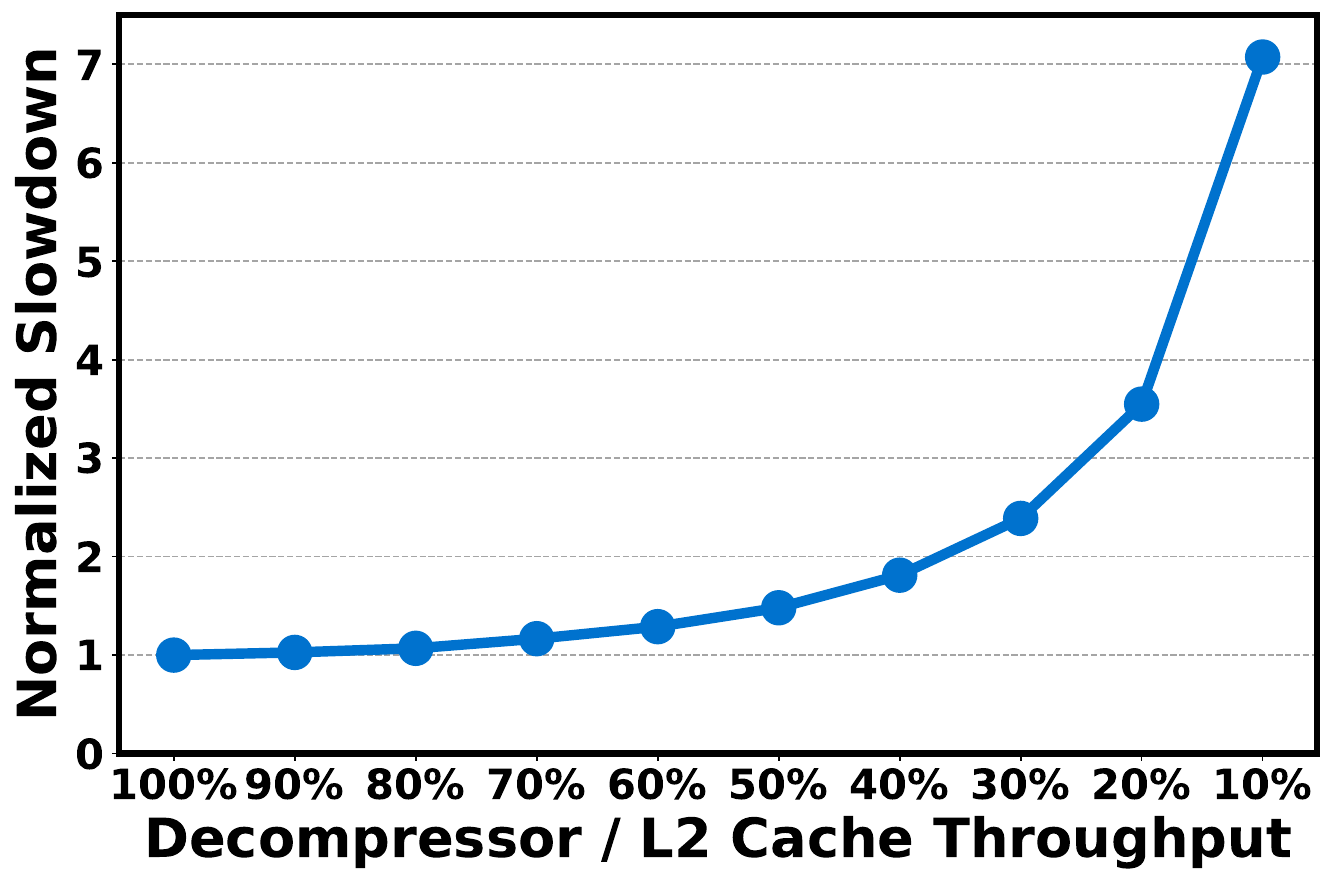}
        \vspace{-1.8em}
        % \caption{Normalized slowdown vs. percentage of L2 peak bandwidth achieved by the decompressor}
        \caption{Slowdown vs. decompressor / L2 Throughput}
        \label{fig:slowdown_bw}
    \end{subfigure}
    \hfill
    \begin{subfigure}[b]{0.48\linewidth}
        \centering
        \includegraphics[width=\linewidth]{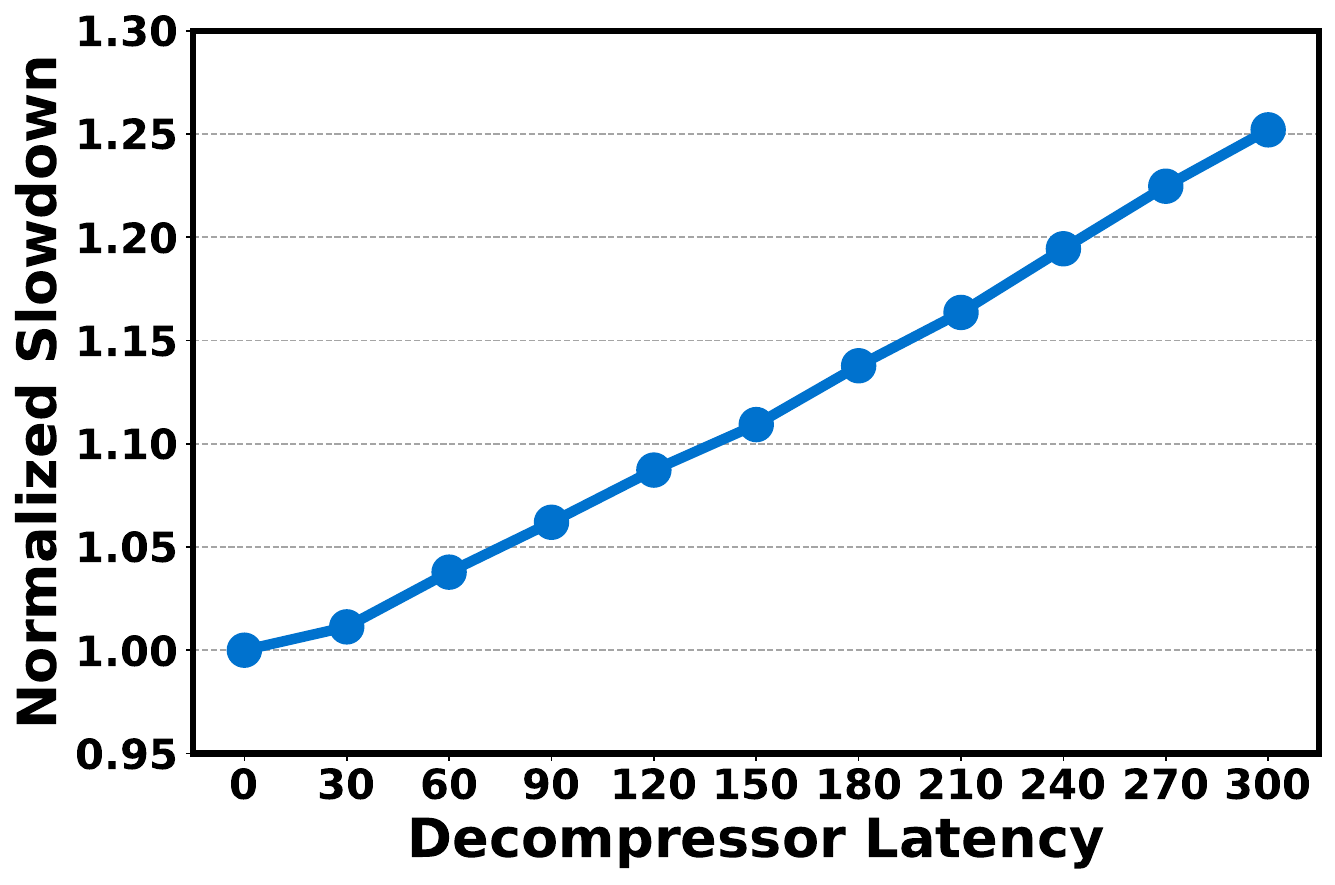}
        \vspace{-1.8em}
        % \caption{\small Normalized slowdown vs.  latency introduced by decompressor}
        \caption{Slowdown vs. decompressor latency}
        \label{fig:slowdown_lat}
    \end{subfigure}
    \vskip 0.8em
    \caption{Sensitivity Analysis: Sweep Decompressor Throughput and Latency.}
    % \vspace{-1em}
    \label{fig:slowdown_bw_lat}
    \Description{Sensitivity Analysis: Sweep Decompressor Throughput and Latency.}
\end{figure}

\noindent \textbf{Sensitivity Analysis.} 
Figure \ref{fig:slowdown_bw_lat} presents a study of how variations in decompressor throughput and latency affect overall performance, measured here as normalized slowdown. In Figure \ref{fig:slowdown_bw}, decompressor throughput is expressed as a fraction of the L2 cache’s peak bandwidth. When the decompressor operates at throughput levels comparable to the L2 bandwidth (90–100\%), the normalized slowdown remains close to one, indicating minimal performance degradation. As throughput declines below about half of the L2 peak, the slowdown starts to grow noticeably, reaching a pronounced increase at 20–10\% of the L2 bandwidth. These results show that although modest reductions in decompressor throughput can be tolerated with little impact on system performance, severe throughput constraints quickly degrade execution time.

Figure \ref{fig:slowdown_lat} illustrates the sensitivity to decompressor latency, measured in cycles. Increasing the decompressor’s latency from near-zero up to hundreds of cycles gradually but steadily raises the normalized slowdown, from just 1.0 to around 1.3. Longer latencies increase the time required to retrieve usable data from compressed blocks, causing more pipeline stalls or waiting periods. Together, these results underscore the importance of designing a decompressor whose throughput and latency closely match—or at least do not lag far behind—those of the cache hierarchy in order to avoid disproportionate penalties to overall performance.

\section{Discussion}

\subsection{\normalsize{Broadening Applicability to CPUs and Accelerators}}

We primarily explored implementing \methodname on GPUs, the most widely used platform for LLM workloads. 
However, our solution can be seamlessly integrated with other platforms as well. 
Accelerators \cite{tpuv6e, cerab, wei2025prosperity}, such as Google TPUs, also suffer from memory-bound issues similar to those observed on GPUs, and we believe \methodname can provide substantial speedups in those contexts. 
In addition, accelerators with relatively small L2 caches may benefit even more, as the compressed data format allows for more data to be stored in the cache without frequent swapping. 
Furthermore, our method is compatible with modern CPUs equipped with AI capabilities, such as Intel Core Ultra Processors \cite{coreultra}, as small-batch computations are memory-bound even on these CPUs. 
Overall, our solution shows promising generalizability across different hardware platforms.
\begin{table}[h]
\centering
\caption{Zero-shot accuracy of LLaMA-3.1-8B-Instruct.}
\footnotesize
\begin{tabular}{l|c|cc|cc}
\toprule

  Model &
\multicolumn{5}{c}{LLaMA-3.1-8B-Instruct}  \\
\midrule
Compression  
&   Ori.    &   \multicolumn{2}{c|}{Weight Only}   &   \multicolumn{2}{c}{Wt. \& Act. \& KV}     \\
\midrule
Method     
& FP16   & AWQ     & \methodname   & QoQ   & \methodname   \\
ARC-c Accuracy
& 83.70 & 81.06  & \textbf{82.85} & 82.17 & \textbf{82.68}\\

\bottomrule

\end{tabular}
\label{tab:evaluation:ft}
\end{table}

\subsection{{From Fine-tuned LLMs to HPC Impact}}

To examine whether \methodname works across a wider range of workloads, we evaluated its performance on the fine-tuned model LLaMA-3.1-Instruct-8B. 
Table \ref{tab:evaluation:ft} shows that our method delivers superior performance compared to other baselines, demonstrating its adaptability.
In addition to fine-tuned LLMs, our method can theoretically be applied to all machine learning workloads, as \methodname is a technique applicable to any matrix multiplication operation. 
Furthermore, it is also suitable for HPC \cite{hpc1, hpc2, hpc3, hpc4} workloads; 
if the compressed data length exceeds the target ratio, our approach performs parallel decompression when accessing the data, whereas the data remains uncompressed if not. 
This underscore the versatility and broad applicability of \methodname across diverse computational workloads.

\section{Conclusion}
In this work, we proposed an entropy-based cache compression method for LLMs, combining group-wise non-uniform quantization with optimized Huffman coding to leverage cache entropy for efficient compression and decompression. Our parallel Huffman-based decoding process and runtime-optimized encoder significantly reduce latency and enhance throughput, seamlessly integrating with traditional cache architectures. This approach achieves up to 2.9\x speedup over AWQ and nearly 4\x memory capacity enlargement, all while maintaining state-of-the-art model accuracy.
These results demonstrate the transformative potential of entropy-based cache compression in overcoming LLM memory bottlenecks. By addressing limitations of existing methods with a scalable, efficient solution, our work enhances LLM accessibility across diverse platforms and lays the foundation for extending such techniques to other memory-intensive AI applications.

\begin{acks}
The work was funded in part by National Science Foundation NSF-2112562 and ARO W911NF-23-2-0224. AMD, the AMD Arrow logo and combinations thereof are trademarks of Advanced Micro Devices, Inc. The authors sincerely thank the anonymous reviewers for their constructive feedback and valuable suggestions that greatly improved the quality of this work.
\end{acks}

%%%%%%% -- PAPER CONTENT ENDS -- %%%%%%%%

%%%%%%%%% -- BIB STYLE AND FILE -- %%%%%%%%
\balance
\bibliographystyle{ACM-Reference-Format}
\bibliography{main}


\begin{thebibliography}{73}

%%% the Web address.

\ifx \showCODEN    \undefined \def \showCODEN     #1{\unskip}     \fi
\ifx \showDOI      \undefined \def \showDOI       #1{#1}\fi
\ifx \showISBNx    \undefined \def \showISBNx     #1{\unskip}     \fi
\ifx \showISBNxiii \undefined \def \showISBNxiii  #1{\unskip}     \fi
\ifx \showISSN     \undefined \def \showISSN      #1{\unskip}     \fi
\ifx \showLCCN     \undefined \def \showLCCN      #1{\unskip}     \fi
\ifx \shownote     \undefined \def \shownote      #1{#1}          \fi
\ifx \showarticletitle \undefined \def \showarticletitle #1{#1}   \fi
\ifx \showURL      \undefined \def \showURL       {\relax}        \fi
% invisible to TeX
\providecommand\bibfield[2]{#2}
\providecommand\bibinfo[2]{#2}
\providecommand\natexlab[1]{#1}
\providecommand\showeprint[2][]{arXiv:#2}

\bibitem[{Advanced Micro Devices}(2023)]%
        {amd2023mi300x}
\bibfield{author}{\bibinfo{person}{{Advanced Micro Devices}}.} \bibinfo{year}{2023}\natexlab{}.
\newblock \bibinfo{booktitle}{\emph{AMD Instinct™ MI300X Accelerator: Technical Overview}}.
\newblock \bibinfo{type}{{T}echnical {R}eport}. \bibinfo{institution}{AMD}.
\newblock
\newblock
\shownote{Product Documentation}.


\bibitem[Alameldeen and Wood(2004)]%
        {alameldeen2004adaptive}
\bibfield{author}{\bibinfo{person}{Alaa~R Alameldeen} {and} \bibinfo{person}{David~A Wood}.} \bibinfo{year}{2004}\natexlab{}.
\newblock \showarticletitle{Adaptive cache compression for high-performance processors}.
\newblock \bibinfo{journal}{\emph{ACM SIGARCH Computer Architecture News}} \bibinfo{volume}{32}, \bibinfo{number}{2} (\bibinfo{year}{2004}), \bibinfo{pages}{212}.
\newblock


\bibitem[Ashkboos et~al\mbox{.}(2024)]%
        {quarot}
\bibfield{author}{\bibinfo{person}{Saleh Ashkboos}, \bibinfo{person}{Amirkeivan Mohtashami}, \bibinfo{person}{Maximilian~L. Croci}, \bibinfo{person}{Bo Li}, \bibinfo{person}{Pashmina Cameron}, \bibinfo{person}{Martin Jaggi}, \bibinfo{person}{Dan Alistarh}, \bibinfo{person}{Torsten Hoefler}, {and} \bibinfo{person}{James Hensman}.} \bibinfo{year}{2024}\natexlab{}.
\newblock \bibinfo{title}{QuaRot: Outlier-Free 4-Bit Inference in Rotated LLMs}.
\newblock
\newblock
\showeprint[arxiv]{2404.00456}~[cs.LG]
\urldef\tempurl%
\url{https://arxiv.org/abs/2404.00456}
\showURL{%
\tempurl}


\bibitem[Bisk et~al\mbox{.}(2019)]%
        {piqa}
\bibfield{author}{\bibinfo{person}{Yonatan Bisk}, \bibinfo{person}{Rowan Zellers}, \bibinfo{person}{Ronan~Le Bras}, \bibinfo{person}{Jianfeng Gao}, {and} \bibinfo{person}{Yejin Choi}.} \bibinfo{year}{2019}\natexlab{}.
\newblock \bibinfo{title}{PIQA: Reasoning about Physical Commonsense in Natural Language}.
\newblock
\newblock
\showeprint[arxiv]{1911.11641}~[cs.CL]
\urldef\tempurl%
\url{https://arxiv.org/abs/1911.11641}
\showURL{%
\tempurl}


\bibitem[Brown et~al\mbox{.}(2020)]%
        {brown2020language}
\bibfield{author}{\bibinfo{person}{Tom Brown}, \bibinfo{person}{Benjamin Mann}, \bibinfo{person}{Nick Ryder}, \bibinfo{person}{Melanie Subbiah}, \bibinfo{person}{Jared~D Kaplan}, \bibinfo{person}{Prafulla Dhariwal}, \bibinfo{person}{Arvind Neelakantan}, \bibinfo{person}{Pranav Shyam}, \bibinfo{person}{Girish Sastry}, \bibinfo{person}{Amanda Askell}, {et~al\mbox{.}}} \bibinfo{year}{2020}\natexlab{}.
\newblock \showarticletitle{Language models are few-shot learners}.
\newblock \bibinfo{journal}{\emph{Advances in neural information processing systems}}  \bibinfo{volume}{33} (\bibinfo{year}{2020}), \bibinfo{pages}{1877--1901}.
\newblock


\bibitem[Chen et~al\mbox{.}(2022)]%
        {hpc4}
\bibfield{author}{\bibinfo{person}{Xinyu Chen}, \bibinfo{person}{Yao Chen}, \bibinfo{person}{Feng Cheng}, \bibinfo{person}{Hongshi Tan}, \bibinfo{person}{Bingsheng He}, {and} \bibinfo{person}{Weng-Fai Wong}.} \bibinfo{year}{2022}\natexlab{}.
\newblock \showarticletitle{ReGraph: Scaling graph processing on HBM-enabled FPGAs with heterogeneous pipelines}. In \bibinfo{booktitle}{\emph{2022 55th IEEE/ACM International Symposium on Microarchitecture (MICRO)}}. IEEE, \bibinfo{pages}{1342--1358}.
\newblock


\bibitem[Choquette et~al\mbox{.}(2021)]%
        {a100}
\bibfield{author}{\bibinfo{person}{Jack Choquette}, \bibinfo{person}{Wishwesh Gandhi}, \bibinfo{person}{Olivier Giroux}, \bibinfo{person}{Nick Stam}, {and} \bibinfo{person}{Ronny Krashinsky}.} \bibinfo{year}{2021}\natexlab{}.
\newblock \showarticletitle{NVIDIA A100 Tensor Core GPU: Performance and Innovation}.
\newblock \bibinfo{journal}{\emph{IEEE Micro}} \bibinfo{volume}{41}, \bibinfo{number}{2} (\bibinfo{year}{2021}), \bibinfo{pages}{29--35}.
\newblock
\urldef\tempurl%
\url{https://doi.org/10.1109/MM.2021.3061394}
\showDOI{\tempurl}


\bibitem[Choukse et~al\mbox{.}(2019)]%
        {buddy}
\bibfield{author}{\bibinfo{person}{Esha Choukse}, \bibinfo{person}{Michael Sullivan}, \bibinfo{person}{Mike O'Connor}, \bibinfo{person}{Mattan Erez}, \bibinfo{person}{Jeff Pool}, \bibinfo{person}{David Nellans}, {and} \bibinfo{person}{Steve Keckler}.} \bibinfo{year}{2019}\natexlab{}.
\newblock \bibinfo{title}{Buddy Compression: Enabling Larger Memory for Deep Learning and HPC Workloads on GPUs}.
\newblock
\newblock
\showeprint[arxiv]{1903.02596}~[cs.AR]
\urldef\tempurl%
\url{https://arxiv.org/abs/1903.02596}
\showURL{%
\tempurl}


\bibitem[Clark et~al\mbox{.}(2018)]%
        {arc}
\bibfield{author}{\bibinfo{person}{Peter Clark}, \bibinfo{person}{Isaac Cowhey}, \bibinfo{person}{Oren Etzioni}, \bibinfo{person}{Tushar Khot}, \bibinfo{person}{Ashish Sabharwal}, \bibinfo{person}{Carissa Schoenick}, {and} \bibinfo{person}{Oyvind Tafjord}.} \bibinfo{year}{2018}\natexlab{}.
\newblock \bibinfo{title}{Think you have Solved Question Answering? Try ARC, the AI2 Reasoning Challenge}.
\newblock
\newblock
\showeprint[arxiv]{1803.05457}~[cs.AI]
\urldef\tempurl%
\url{https://arxiv.org/abs/1803.05457}
\showURL{%
\tempurl}


\bibitem[Dettmers et~al\mbox{.}(2023)]%
        {dettmers2024qlora}
\bibfield{author}{\bibinfo{person}{Tim Dettmers}, \bibinfo{person}{Artidoro Pagnoni}, \bibinfo{person}{Ari Holtzman}, {and} \bibinfo{person}{Luke Zettlemoyer}.} \bibinfo{year}{2023}\natexlab{}.
\newblock \bibinfo{title}{QLoRA: Efficient Finetuning of Quantized LLMs}.
\newblock
\newblock
\showeprint[arxiv]{2305.14314}~[cs.LG]
\urldef\tempurl%
\url{https://arxiv.org/abs/2305.14314}
\showURL{%
\tempurl}


\bibitem[Evans et~al\mbox{.}(2020)]%
        {jpeg}
\bibfield{author}{\bibinfo{person}{R.~David Evans}, \bibinfo{person}{Lufei Liu}, {and} \bibinfo{person}{Tor~M. Aamodt}.} \bibinfo{year}{2020}\natexlab{}.
\newblock \showarticletitle{JPEG-ACT: accelerating deep learning via transform-based lossy compression}. In \bibinfo{booktitle}{\emph{Proceedings of the ACM/IEEE 47th Annual International Symposium on Computer Architecture}} \emph{(\bibinfo{series}{ISCA '20})}. \bibinfo{publisher}{IEEE Press}, \bibinfo{address}{Virtual Event}, \bibinfo{pages}{860–873}.
\newblock
\showISBNx{9781728146614}
\urldef\tempurl%
\url{https://doi.org/10.1109/ISCA45697.2020.00075}
\showDOI{\tempurl}


\bibitem[Frantar et~al\mbox{.}(2023)]%
        {gptq}
\bibfield{author}{\bibinfo{person}{Elias Frantar}, \bibinfo{person}{Saleh Ashkboos}, \bibinfo{person}{Torsten Hoefler}, {and} \bibinfo{person}{Dan Alistarh}.} \bibinfo{year}{2023}\natexlab{}.
\newblock \bibinfo{title}{GPTQ: Accurate Post-Training Quantization for Generative Pre-trained Transformers}.
\newblock
\newblock
\showeprint[arxiv]{2210.17323}~[cs.LG]
\urldef\tempurl%
\url{https://arxiv.org/abs/2210.17323}
\showURL{%
\tempurl}


\bibitem[Gao et~al\mbox{.}(2020)]%
        {pileval}
\bibfield{author}{\bibinfo{person}{Leo Gao}, \bibinfo{person}{Stella Biderman}, \bibinfo{person}{Sid Black}, \bibinfo{person}{Laurence Golding}, \bibinfo{person}{Travis Hoppe}, \bibinfo{person}{Charles Foster}, \bibinfo{person}{Jason Phang}, \bibinfo{person}{Horace He}, \bibinfo{person}{Anish Thite}, \bibinfo{person}{Noa Nabeshima}, \bibinfo{person}{Shawn Presser}, {and} \bibinfo{person}{Connor Leahy}.} \bibinfo{year}{2020}\natexlab{}.
\newblock \bibinfo{title}{The Pile: An 800GB Dataset of Diverse Text for Language Modeling}.
\newblock
\newblock
\showeprint[arxiv]{2101.00027}~[cs.CL]
\urldef\tempurl%
\url{https://arxiv.org/abs/2101.00027}
\showURL{%
\tempurl}


\bibitem[Gao et~al\mbox{.}(2024a)]%
        {eval-harness}
\bibfield{author}{\bibinfo{person}{Leo Gao}, \bibinfo{person}{Jonathan Tow}, \bibinfo{person}{Baber Abbasi}, \bibinfo{person}{Stella Biderman}, \bibinfo{person}{Sid Black}, \bibinfo{person}{Anthony DiPofi}, \bibinfo{person}{Charles Foster}, \bibinfo{person}{Laurence Golding}, \bibinfo{person}{Jeffrey Hsu}, \bibinfo{person}{Alain Le~Noac'h}, \bibinfo{person}{Haonan Li}, \bibinfo{person}{Kyle McDonell}, \bibinfo{person}{Niklas Muennighoff}, \bibinfo{person}{Chris Ociepa}, \bibinfo{person}{Jason Phang}, \bibinfo{person}{Laria Reynolds}, \bibinfo{person}{Hailey Schoelkopf}, \bibinfo{person}{Aviya Skowron}, \bibinfo{person}{Lintang Sutawika}, \bibinfo{person}{Eric Tang}, \bibinfo{person}{Anish Thite}, \bibinfo{person}{Ben Wang}, \bibinfo{person}{Kevin Wang}, {and} \bibinfo{person}{Andy Zou}.} \bibinfo{year}{2024}\natexlab{a}.
\newblock \bibinfo{title}{A framework for few-shot language model evaluation}.
\newblock
\newblock
\urldef\tempurl%
\url{https://doi.org/10.5281/zenodo.12608602}
\showDOI{\tempurl}


\bibitem[Gao et~al\mbox{.}(2024b)]%
        {llm_gen}
\bibfield{author}{\bibinfo{person}{Yunfan Gao}, \bibinfo{person}{Yun Xiong}, \bibinfo{person}{Xinyu Gao}, \bibinfo{person}{Kangxiang Jia}, \bibinfo{person}{Jinliu Pan}, \bibinfo{person}{Yuxi Bi}, \bibinfo{person}{Yi Dai}, \bibinfo{person}{Jiawei Sun}, \bibinfo{person}{Meng Wang}, {and} \bibinfo{person}{Haofen Wang}.} \bibinfo{year}{2024}\natexlab{b}.
\newblock \bibinfo{title}{Retrieval-Augmented Generation for Large Language Models: A Survey}.
\newblock
\newblock
\showeprint[arxiv]{2312.10997}~[cs.CL]
\urldef\tempurl%
\url{https://arxiv.org/abs/2312.10997}
\showURL{%
\tempurl}


\bibitem[Google(2024)]%
        {tpuv6e}
\bibfield{author}{\bibinfo{person}{Google}.} \bibinfo{year}{2024}\natexlab{}.
\newblock \bibinfo{title}{TPU V6e}.
\newblock
\newblock
\urldef\tempurl%
\url{https://cloud.google.com/blog/products/compute/introducing-trillium-6th-gen-tpus}
\showURL{%
\tempurl}


\bibitem[Guo et~al\mbox{.}(2025a)]%
        {guo2025survey}
\bibfield{author}{\bibinfo{person}{Cong Guo}, \bibinfo{person}{Feng Cheng}, \bibinfo{person}{Zhixu Du}, \bibinfo{person}{James Kiessling}, \bibinfo{person}{Jonathan Ku}, \bibinfo{person}{Shiyu Li}, \bibinfo{person}{Ziru Li}, \bibinfo{person}{Mingyuan Ma}, \bibinfo{person}{Tergel Molom-Ochir}, \bibinfo{person}{Benjamin Morris}, {et~al\mbox{.}}} \bibinfo{year}{2025}\natexlab{a}.
\newblock \showarticletitle{A Survey: Collaborative Hardware and Software Design in the Era of Large Language Models}.
\newblock \bibinfo{journal}{\emph{IEEE Circuits and Systems Magazine}} \bibinfo{volume}{25}, \bibinfo{number}{1} (\bibinfo{year}{2025}), \bibinfo{pages}{35--57}.
\newblock


\bibitem[Guo et~al\mbox{.}(2025b)]%
        {llmsurvey}
\bibfield{author}{\bibinfo{person}{Cong Guo}, \bibinfo{person}{Feng Cheng}, \bibinfo{person}{Zhixu Du}, \bibinfo{person}{James Kiessling}, \bibinfo{person}{Jonathan Ku}, \bibinfo{person}{Shiyu Li}, \bibinfo{person}{Ziru Li}, \bibinfo{person}{Mingyuan Ma}, \bibinfo{person}{Tergel Molom-Ochir}, \bibinfo{person}{Benjamin Morris}, \bibinfo{person}{Haoxuan Shan}, \bibinfo{person}{Jingwei Sun}, \bibinfo{person}{Yitu Wang}, \bibinfo{person}{Chiyue Wei}, \bibinfo{person}{Xueying Wu}, \bibinfo{person}{Yuhao Wu}, \bibinfo{person}{Hao~Frank Yang}, \bibinfo{person}{Jingyang Zhang}, \bibinfo{person}{Junyao Zhang}, \bibinfo{person}{Qilin Zheng}, \bibinfo{person}{Guanglei Zhou}, \bibinfo{person}{Hai Li}, {and} \bibinfo{person}{Yiran Chen}.} \bibinfo{year}{2025}\natexlab{b}.
\newblock \showarticletitle{A Survey: Collaborative Hardware and Software Design in the Era of Large Language Models}.
\newblock \bibinfo{journal}{\emph{IEEE Circuits and Systems Magazine}} \bibinfo{volume}{25}, \bibinfo{number}{1} (\bibinfo{year}{2025}), \bibinfo{pages}{35--57}.
\newblock
\urldef\tempurl%
\url{https://doi.org/10.1109/MCAS.2024.3476008}
\showDOI{\tempurl}


\bibitem[Guo et~al\mbox{.}(2020)]%
        {guo2020accelerating}
\bibfield{author}{\bibinfo{person}{Cong Guo}, \bibinfo{person}{Bo~Yang Hsueh}, \bibinfo{person}{Jingwen Leng}, \bibinfo{person}{Yuxian Qiu}, \bibinfo{person}{Yue Guan}, \bibinfo{person}{Zehuan Wang}, \bibinfo{person}{Xiaoying Jia}, \bibinfo{person}{Xipeng Li}, \bibinfo{person}{Minyi Guo}, {and} \bibinfo{person}{Yuhao Zhu}.} \bibinfo{year}{2020}\natexlab{}.
\newblock \showarticletitle{Accelerating sparse dnn models without hardware-support via tile-wise sparsity}. In \bibinfo{booktitle}{\emph{SC20: International Conference for High Performance Computing, Networking, Storage and Analysis}}. IEEE, \bibinfo{pages}{1--15}.
\newblock


\bibitem[Guo et~al\mbox{.}(2022a)]%
        {guo2022squant}
\bibfield{author}{\bibinfo{person}{Cong Guo}, \bibinfo{person}{Yuxian Qiu}, \bibinfo{person}{Jingwen Leng}, \bibinfo{person}{Xiaotian Gao}, \bibinfo{person}{Chen Zhang}, \bibinfo{person}{Yunxin Liu}, \bibinfo{person}{Fan Yang}, \bibinfo{person}{Yuhao Zhu}, {and} \bibinfo{person}{Minyi Guo}.} \bibinfo{year}{2022}\natexlab{a}.
\newblock \showarticletitle{{SQ}uant: On-the-Fly Data-Free Quantization via Diagonal Hessian Approximation}. In \bibinfo{booktitle}{\emph{International Conference on Learning Representations}}.
\newblock
\urldef\tempurl%
\url{https://openreview.net/forum?id=JXhROKNZzOc}
\showURL{%
\tempurl}


\bibitem[Guo et~al\mbox{.}(2023)]%
        {olive}
\bibfield{author}{\bibinfo{person}{Cong Guo}, \bibinfo{person}{Jiaming Tang}, \bibinfo{person}{Weiming Hu}, \bibinfo{person}{Jingwen Leng}, \bibinfo{person}{Chen Zhang}, \bibinfo{person}{Fan Yang}, \bibinfo{person}{Yunxin Liu}, \bibinfo{person}{Minyi Guo}, {and} \bibinfo{person}{Yuhao Zhu}.} \bibinfo{year}{2023}\natexlab{}.
\newblock \showarticletitle{OliVe: Accelerating Large Language Models via Hardware-friendly Outlier-Victim Pair Quantization}. In \bibinfo{booktitle}{\emph{Proceedings of the 50th Annual International Symposium on Computer Architecture}} (Orlando, FL, USA) \emph{(\bibinfo{series}{ISCA '23})}. \bibinfo{publisher}{Association for Computing Machinery}, \bibinfo{address}{New York, NY, USA}, Article \bibinfo{articleno}{3}, \bibinfo{numpages}{15}~pages.
\newblock
\showISBNx{9798400700958}
\urldef\tempurl%
\url{https://doi.org/10.1145/3579371.3589038}
\showDOI{\tempurl}


\bibitem[Guo et~al\mbox{.}(2024)]%
        {guo2024accelerating}
\bibfield{author}{\bibinfo{person}{Cong Guo}, \bibinfo{person}{Fengchen Xue}, \bibinfo{person}{Jingwen Leng}, \bibinfo{person}{Yuxian Qiu}, \bibinfo{person}{Yue Guan}, \bibinfo{person}{Weihao Cui}, \bibinfo{person}{Quan Chen}, {and} \bibinfo{person}{Minyi Guo}.} \bibinfo{year}{2024}\natexlab{}.
\newblock \showarticletitle{Accelerating sparse dnns based on tiled gemm}.
\newblock \bibinfo{journal}{\emph{IEEE Trans. Comput.}} (\bibinfo{year}{2024}).
\newblock


\bibitem[Guo et~al\mbox{.}(2022b)]%
        {guo2022ant}
\bibfield{author}{\bibinfo{person}{Cong Guo}, \bibinfo{person}{Chen Zhang}, \bibinfo{person}{Jingwen Leng}, \bibinfo{person}{Zihan Liu}, \bibinfo{person}{Fan Yang}, \bibinfo{person}{Yunxin Liu}, \bibinfo{person}{Minyi Guo}, {and} \bibinfo{person}{Yuhao Zhu}.} \bibinfo{year}{2022}\natexlab{b}.
\newblock \showarticletitle{Ant: Exploiting adaptive numerical data type for low-bit deep neural network quantization}. In \bibinfo{booktitle}{\emph{2022 55th IEEE/ACM International Symposium on Microarchitecture (MICRO)}}. IEEE, \bibinfo{pages}{1414--1433}.
\newblock


\bibitem[Han et~al\mbox{.}(2016a)]%
        {han2016eie}
\bibfield{author}{\bibinfo{person}{Song Han}, \bibinfo{person}{Xingyu Liu}, \bibinfo{person}{Huizi Mao}, \bibinfo{person}{Jing Pu}, \bibinfo{person}{Ardavan Pedram}, \bibinfo{person}{Mark~A Horowitz}, {and} \bibinfo{person}{William~J Dally}.} \bibinfo{year}{2016}\natexlab{a}.
\newblock \showarticletitle{EIE: Efficient inference engine on compressed deep neural network}.
\newblock \bibinfo{journal}{\emph{ACM SIGARCH Computer Architecture News}} \bibinfo{volume}{44}, \bibinfo{number}{3} (\bibinfo{year}{2016}), \bibinfo{pages}{243--254}.
\newblock


\bibitem[Han et~al\mbox{.}(2016b)]%
        {han2015deep}
\bibfield{author}{\bibinfo{person}{Song Han}, \bibinfo{person}{Huizi Mao}, {and} \bibinfo{person}{William~J. Dally}.} \bibinfo{year}{2016}\natexlab{b}.
\newblock \bibinfo{title}{Deep Compression: Compressing Deep Neural Networks with Pruning, Trained Quantization and Huffman Coding}.
\newblock
\newblock
\showeprint[arxiv]{1510.00149}~[cs.CV]
\urldef\tempurl%
\url{https://arxiv.org/abs/1510.00149}
\showURL{%
\tempurl}


\bibitem[Hanson et~al\mbox{.}(2023)]%
        {cerab}
\bibfield{author}{\bibinfo{person}{Edward Hanson}, \bibinfo{person}{Shiyu Li}, \bibinfo{person}{Guanglei Zhou}, \bibinfo{person}{Feng Cheng}, \bibinfo{person}{Yitu Wang}, \bibinfo{person}{Rohan Bose}, \bibinfo{person}{Hai Li}, {and} \bibinfo{person}{Yiran Chen}.} \bibinfo{year}{2023}\natexlab{}.
\newblock \showarticletitle{Si-kintsugi: Towards recovering golden-like performance of defective many-core spatial architectures for ai}. In \bibinfo{booktitle}{\emph{Proceedings of the 56th Annual IEEE/ACM International Symposium on Microarchitecture}}. \bibinfo{pages}{972--985}.
\newblock


\bibitem[Hooper et~al\mbox{.}(2024)]%
        {hooper2024kvquant}
\bibfield{author}{\bibinfo{person}{Coleman Hooper}, \bibinfo{person}{Sehoon Kim}, \bibinfo{person}{Hiva Mohammadzadeh}, \bibinfo{person}{Michael~W. Mahoney}, \bibinfo{person}{Yakun~Sophia Shao}, \bibinfo{person}{Kurt Keutzer}, {and} \bibinfo{person}{Amir Gholami}.} \bibinfo{year}{2024}\natexlab{}.
\newblock \bibinfo{title}{KVQuant: Towards 10 Million Context Length LLM Inference with KV Cache Quantization}.
\newblock
\newblock
\showeprint[arxiv]{2401.18079}~[cs.LG]
\urldef\tempurl%
\url{https://arxiv.org/abs/2401.18079}
\showURL{%
\tempurl}


\bibitem[Hu et~al\mbox{.}(2025)]%
        {hu2025m}
\bibfield{author}{\bibinfo{person}{Weiming Hu}, \bibinfo{person}{Haoyan Zhang}, \bibinfo{person}{Cong Guo}, \bibinfo{person}{Yu Feng}, \bibinfo{person}{Renyang Guan}, \bibinfo{person}{Zhendong Hua}, \bibinfo{person}{Zihan Liu}, \bibinfo{person}{Yue Guan}, \bibinfo{person}{Minyi Guo}, {and} \bibinfo{person}{Jingwen Leng}.} \bibinfo{year}{2025}\natexlab{}.
\newblock \showarticletitle{M-ANT: Efficient Low-bit Group Quantization for LLMs via Mathematically Adaptive Numerical Type}. In \bibinfo{booktitle}{\emph{2025 IEEE International Symposium on High Performance Computer Architecture (HPCA)}}. IEEE, \bibinfo{pages}{1112--1126}.
\newblock


\bibitem[Huang and Chang(2023)]%
        {llm_reasoning}
\bibfield{author}{\bibinfo{person}{Jie Huang} {and} \bibinfo{person}{Kevin Chen-Chuan Chang}.} \bibinfo{year}{2023}\natexlab{}.
\newblock \bibinfo{title}{Towards Reasoning in Large Language Models: A Survey}.
\newblock
\newblock
\showeprint[arxiv]{2212.10403}~[cs.CL]
\urldef\tempurl%
\url{https://arxiv.org/abs/2212.10403}
\showURL{%
\tempurl}


\bibitem[Huffman(1952)]%
        {huffman1952method}
\bibfield{author}{\bibinfo{person}{David~A Huffman}.} \bibinfo{year}{1952}\natexlab{}.
\newblock \showarticletitle{A method for the construction of minimum-redundancy codes}.
\newblock \bibinfo{journal}{\emph{Proceedings of the IRE}} \bibinfo{volume}{40}, \bibinfo{number}{9} (\bibinfo{year}{1952}), \bibinfo{pages}{1098--1101}.
\newblock


\bibitem[Intel(2024)]%
        {coreultra}
\bibfield{author}{\bibinfo{person}{Intel}.} \bibinfo{year}{2024}\natexlab{}.
\newblock \bibinfo{title}{Intel Core Ultra}.
\newblock
\newblock
\urldef\tempurl%
\url{https://www.intel.com/content/www/us/en/products/sku/241747/intel-core-ultra-9-processor-285h-24m-cache-up-to-5-40-ghz/specifications.html}
\showURL{%
\tempurl}


\bibitem[Jacob et~al\mbox{.}(2017)]%
        {jacob2018quantization}
\bibfield{author}{\bibinfo{person}{Benoit Jacob}, \bibinfo{person}{Skirmantas Kligys}, \bibinfo{person}{Bo Chen}, \bibinfo{person}{Menglong Zhu}, \bibinfo{person}{Matthew Tang}, \bibinfo{person}{Andrew Howard}, \bibinfo{person}{Hartwig Adam}, {and} \bibinfo{person}{Dmitry Kalenichenko}.} \bibinfo{year}{2017}\natexlab{}.
\newblock \bibinfo{title}{Quantization and Training of Neural Networks for Efficient Integer-Arithmetic-Only Inference}.
\newblock
\newblock
\showeprint[arxiv]{1712.05877}~[cs.LG]
\urldef\tempurl%
\url{https://arxiv.org/abs/1712.05877}
\showURL{%
\tempurl}


\bibitem[Jain et~al\mbox{.}(2018)]%
        {jain2018gist}
\bibfield{author}{\bibinfo{person}{Animesh Jain}, \bibinfo{person}{Amar Phanishayee}, \bibinfo{person}{Jason Mars}, \bibinfo{person}{Lingjia Tang}, {and} \bibinfo{person}{Gennady Pekhimenko}.} \bibinfo{year}{2018}\natexlab{}.
\newblock \showarticletitle{Gist: efficient data encoding for deep neural network training}. In \bibinfo{booktitle}{\emph{Proceedings of the 45th Annual International Symposium on Computer Architecture}} \emph{(\bibinfo{series}{ISCA '18})}. \bibinfo{publisher}{IEEE Press}, \bibinfo{address}{Los Angeles, California}, \bibinfo{pages}{776–789}.
\newblock
\showISBNx{9781538659847}
\urldef\tempurl%
\url{https://doi.org/10.1109/ISCA.2018.00070}
\showDOI{\tempurl}


\bibitem[Jiang et~al\mbox{.}(2024)]%
        {mistral}
\bibfield{author}{\bibinfo{person}{Albert~Q. Jiang}, \bibinfo{person}{Alexandre Sablayrolles}, \bibinfo{person}{Antoine Roux}, \bibinfo{person}{Arthur Mensch}, \bibinfo{person}{Blanche Savary}, \bibinfo{person}{Chris Bamford}, \bibinfo{person}{Devendra~Singh Chaplot}, \bibinfo{person}{Diego de~las Casas}, \bibinfo{person}{Emma~Bou Hanna}, \bibinfo{person}{Florian Bressand}, \bibinfo{person}{Gianna Lengyel}, \bibinfo{person}{Guillaume Bour}, \bibinfo{person}{Guillaume Lample}, \bibinfo{person}{Lélio~Renard Lavaud}, \bibinfo{person}{Lucile Saulnier}, \bibinfo{person}{Marie-Anne Lachaux}, \bibinfo{person}{Pierre Stock}, \bibinfo{person}{Sandeep Subramanian}, \bibinfo{person}{Sophia Yang}, \bibinfo{person}{Szymon Antoniak}, \bibinfo{person}{Teven~Le Scao}, \bibinfo{person}{Théophile Gervet}, \bibinfo{person}{Thibaut Lavril}, \bibinfo{person}{Thomas Wang}, \bibinfo{person}{Timothée Lacroix}, {and} \bibinfo{person}{William~El Sayed}.} \bibinfo{year}{2024}\natexlab{}.
\newblock \bibinfo{title}{Mixtral of Experts}.
\newblock
\newblock
\showeprint[arxiv]{2401.04088}~[cs.LG]
\urldef\tempurl%
\url{https://arxiv.org/abs/2401.04088}
\showURL{%
\tempurl}


\bibitem[Kaplan et~al\mbox{.}(2020)]%
        {kaplan2020scaling}
\bibfield{author}{\bibinfo{person}{Jared Kaplan}, \bibinfo{person}{Sam McCandlish}, \bibinfo{person}{Tom Henighan}, \bibinfo{person}{Tom~B. Brown}, \bibinfo{person}{Benjamin Chess}, \bibinfo{person}{Rewon Child}, \bibinfo{person}{Scott Gray}, \bibinfo{person}{Alec Radford}, \bibinfo{person}{Jeffrey Wu}, {and} \bibinfo{person}{Dario Amodei}.} \bibinfo{year}{2020}\natexlab{}.
\newblock \bibinfo{title}{Scaling Laws for Neural Language Models}.
\newblock
\newblock
\showeprint[arxiv]{2001.08361}~[cs.LG]
\urldef\tempurl%
\url{https://arxiv.org/abs/2001.08361}
\showURL{%
\tempurl}


\bibitem[Khairy et~al\mbox{.}(2020)]%
        {accelsim}
\bibfield{author}{\bibinfo{person}{Mahmoud Khairy}, \bibinfo{person}{Zhesheng Shen}, \bibinfo{person}{Tor~M. Aamodt}, {and} \bibinfo{person}{Timothy~G. Rogers}.} \bibinfo{year}{2020}\natexlab{}.
\newblock \showarticletitle{Accel-sim: an extensible simulation framework for validated GPU modeling}. In \bibinfo{booktitle}{\emph{Proceedings of the ACM/IEEE 47th Annual International Symposium on Computer Architecture}} \emph{(\bibinfo{series}{ISCA '20})}. \bibinfo{publisher}{IEEE Press}, \bibinfo{address}{Virtual Event}, \bibinfo{pages}{473–486}.
\newblock
\showISBNx{9781728146614}
\urldef\tempurl%
\url{https://doi.org/10.1109/ISCA45697.2020.00047}
\showDOI{\tempurl}


\bibitem[Kim et~al\mbox{.}(2024)]%
        {kim2023squeezellm}
\bibfield{author}{\bibinfo{person}{Sehoon Kim}, \bibinfo{person}{Coleman Hooper}, \bibinfo{person}{Amir Gholami}, \bibinfo{person}{Zhen Dong}, \bibinfo{person}{Xiuyu Li}, \bibinfo{person}{Sheng Shen}, \bibinfo{person}{Michael~W. Mahoney}, {and} \bibinfo{person}{Kurt Keutzer}.} \bibinfo{year}{2024}\natexlab{}.
\newblock \bibinfo{title}{SqueezeLLM: Dense-and-Sparse Quantization}.
\newblock
\newblock
\showeprint[arxiv]{2306.07629}~[cs.CL]
\urldef\tempurl%
\url{https://arxiv.org/abs/2306.07629}
\showURL{%
\tempurl}


\bibitem[Kim et~al\mbox{.}(2022)]%
        {elephant}
\bibfield{author}{\bibinfo{person}{Young~Jin Kim}, \bibinfo{person}{Rawn Henry}, \bibinfo{person}{Raffy Fahim}, {and} \bibinfo{person}{Hany~Hassan Awadalla}.} \bibinfo{year}{2022}\natexlab{}.
\newblock \bibinfo{title}{Who Says Elephants Can't Run: Bringing Large Scale MoE Models into Cloud Scale Production}.
\newblock
\newblock
\showeprint[arxiv]{2211.10017}~[cs.CL]
\urldef\tempurl%
\url{https://arxiv.org/abs/2211.10017}
\showURL{%
\tempurl}


\bibitem[Kwon et~al\mbox{.}(2023)]%
        {vllm}
\bibfield{author}{\bibinfo{person}{Woosuk Kwon}, \bibinfo{person}{Zhuohan Li}, \bibinfo{person}{Siyuan Zhuang}, \bibinfo{person}{Ying Sheng}, \bibinfo{person}{Lianmin Zheng}, \bibinfo{person}{Cody~Hao Yu}, \bibinfo{person}{Joseph Gonzalez}, \bibinfo{person}{Hao Zhang}, {and} \bibinfo{person}{Ion Stoica}.} \bibinfo{year}{2023}\natexlab{}.
\newblock \showarticletitle{Efficient Memory Management for Large Language Model Serving with PagedAttention}. In \bibinfo{booktitle}{\emph{Proceedings of the 29th Symposium on Operating Systems Principles}} (Koblenz, Germany) \emph{(\bibinfo{series}{SOSP '23})}. \bibinfo{publisher}{Association for Computing Machinery}, \bibinfo{address}{New York, NY, USA}, \bibinfo{pages}{611–626}.
\newblock
\showISBNx{9798400702297}
\urldef\tempurl%
\url{https://doi.org/10.1145/3600006.3613165}
\showDOI{\tempurl}


\bibitem[Lew et~al\mbox{.}(2019)]%
        {gpgpusim}
\bibfield{author}{\bibinfo{person}{Jonathan Lew}, \bibinfo{person}{Deval Shah}, \bibinfo{person}{Suchita Pati}, \bibinfo{person}{Shaylin Cattell}, \bibinfo{person}{Mengchi Zhang}, \bibinfo{person}{Amruth Sandhupatla}, \bibinfo{person}{Christopher Ng}, \bibinfo{person}{Negar Goli}, \bibinfo{person}{Matthew~D. Sinclair}, \bibinfo{person}{Timothy~G. Rogers}, {and} \bibinfo{person}{Tor Aamodt}.} \bibinfo{year}{2019}\natexlab{}.
\newblock \bibinfo{title}{Analyzing Machine Learning Workloads Using a Detailed GPU Simulator}.
\newblock
\newblock
\showeprint[arxiv]{1811.08933}~[cs.DC]
\urldef\tempurl%
\url{https://arxiv.org/abs/1811.08933}
\showURL{%
\tempurl}


\bibitem[Li et~al\mbox{.}(2025)]%
        {li2024large}
\bibfield{author}{\bibinfo{person}{Jinhao Li}, \bibinfo{person}{Jiaming Xu}, \bibinfo{person}{Shan Huang}, \bibinfo{person}{Yonghua Chen}, \bibinfo{person}{Wen Li}, \bibinfo{person}{Jun Liu}, \bibinfo{person}{Yaoxiu Lian}, \bibinfo{person}{Jiayi Pan}, \bibinfo{person}{Li Ding}, \bibinfo{person}{Hao Zhou}, \bibinfo{person}{Yu Wang}, {and} \bibinfo{person}{Guohao Dai}.} \bibinfo{year}{2025}\natexlab{}.
\newblock \bibinfo{title}{Large Language Model Inference Acceleration: A Comprehensive Hardware Perspective}.
\newblock
\newblock
\showeprint[arxiv]{2410.04466}~[cs.AR]
\urldef\tempurl%
\url{https://arxiv.org/abs/2410.04466}
\showURL{%
\tempurl}


\bibitem[Lin et~al\mbox{.}(2024a)]%
        {lin2024awq}
\bibfield{author}{\bibinfo{person}{Ji Lin}, \bibinfo{person}{Jiaming Tang}, \bibinfo{person}{Haotian Tang}, \bibinfo{person}{Shang Yang}, \bibinfo{person}{Wei-Ming Chen}, \bibinfo{person}{Wei-Chen Wang}, \bibinfo{person}{Guangxuan Xiao}, \bibinfo{person}{Xingyu Dang}, \bibinfo{person}{Chuang Gan}, {and} \bibinfo{person}{Song Han}.} \bibinfo{year}{2024}\natexlab{a}.
\newblock \showarticletitle{AWQ: Activation-aware Weight Quantization for On-Device LLM Compression and Acceleration}.
\newblock \bibinfo{journal}{\emph{Proceedings of Machine Learning and Systems}}  \bibinfo{volume}{6} (\bibinfo{year}{2024}), \bibinfo{pages}{87--100}.
\newblock


\bibitem[Lin et~al\mbox{.}(2024b)]%
        {lin2024qserve}
\bibfield{author}{\bibinfo{person}{Yujun Lin}, \bibinfo{person}{Haotian Tang}, \bibinfo{person}{Shang Yang}, \bibinfo{person}{Zhekai Zhang}, \bibinfo{person}{Guangxuan Xiao}, \bibinfo{person}{Chuang Gan}, {and} \bibinfo{person}{Song Han}.} \bibinfo{year}{2024}\natexlab{b}.
\newblock \bibinfo{title}{QServe: W4A8KV4 Quantization and System Co-design for Efficient LLM Serving}.
\newblock
\newblock
\showeprint[arxiv]{2405.04532}~[cs.CL]
\urldef\tempurl%
\url{https://arxiv.org/abs/2405.04532}
\showURL{%
\tempurl}


\bibitem[MacQueen(1967)]%
        {macqueen1967some}
\bibfield{author}{\bibinfo{person}{James MacQueen}.} \bibinfo{year}{1967}\natexlab{}.
\newblock \showarticletitle{Some methods for classification and analysis of multivariate observations}. In \bibinfo{booktitle}{\emph{Proceedings of the Fifth Berkeley Symposium on Mathematical Statistics and Probability, Volume 1: Statistics}}, Vol.~\bibinfo{volume}{5}. \bibinfo{publisher}{University of California press}, \bibinfo{address}{Oakland, CA, USA}, \bibinfo{pages}{281--298}.
\newblock


\bibitem[Merity et~al\mbox{.}(2016)]%
        {wiki2}
\bibfield{author}{\bibinfo{person}{Stephen Merity}, \bibinfo{person}{Caiming Xiong}, \bibinfo{person}{James Bradbury}, {and} \bibinfo{person}{Richard Socher}.} \bibinfo{year}{2016}\natexlab{}.
\newblock \bibinfo{title}{Pointer Sentinel Mixture Models}.
\newblock
\newblock
\showeprint[arxiv]{1609.07843}~[cs.CL]
\urldef\tempurl%
\url{https://arxiv.org/abs/1609.07843}
\showURL{%
\tempurl}


\bibitem[Min et~al\mbox{.}(2023)]%
        {llm_nlp}
\bibfield{author}{\bibinfo{person}{Bonan Min}, \bibinfo{person}{Hayley Ross}, \bibinfo{person}{Elior Sulem}, \bibinfo{person}{Amir Pouran~Ben Veyseh}, \bibinfo{person}{Thien~Huu Nguyen}, \bibinfo{person}{Oscar Sainz}, \bibinfo{person}{Eneko Agirre}, \bibinfo{person}{Ilana Heintz}, {and} \bibinfo{person}{Dan Roth}.} \bibinfo{year}{2023}\natexlab{}.
\newblock \showarticletitle{Recent advances in natural language processing via large pre-trained language models: A survey}.
\newblock \bibinfo{journal}{\emph{Comput. Surveys}} \bibinfo{volume}{56}, \bibinfo{number}{2} (\bibinfo{year}{2023}), \bibinfo{pages}{1--40}.
\newblock


\bibitem[NVIDIA(2020)]%
        {gtc}
\bibfield{author}{\bibinfo{person}{NVIDIA}.} \bibinfo{year}{2020}\natexlab{}.
\newblock \bibinfo{title}{GTC 2020}.
\newblock
\newblock
\urldef\tempurl%
\url{https://developer.download.nvidia.cn/video/gputechconf/gtc/2020/presentations/s21819-optimizing-applications-for-nvidia-ampere-gpu-architecture.pdf}
\showURL{%
\tempurl}


\bibitem[NVIDIA(2024a)]%
        {cdc-a100}
\bibfield{author}{\bibinfo{person}{NVIDIA}.} \bibinfo{year}{2024}\natexlab{a}.
\newblock \bibinfo{title}{A100 Compute Data Compression}.
\newblock
\newblock
\urldef\tempurl%
\url{https://hc32.hotchips.org/assets/program/conference/day1/HotChips2020_GPU_NVIDIA_Choquette_v01.pdf#page=22}
\showURL{%
\tempurl}


\bibitem[NVIDIA(2024b)]%
        {compressiblemem}
\bibfield{author}{\bibinfo{person}{NVIDIA}.} \bibinfo{year}{2024}\natexlab{b}.
\newblock \bibinfo{title}{Compressible Memory}.
\newblock
\newblock
\urldef\tempurl%
\url{https://docs.nvidia.com/cuda/cuda-c-programming-guide/index.html#compressible-memory}
\showURL{%
\tempurl}


\bibitem[NVIDIA(2024c)]%
        {cublas}
\bibfield{author}{\bibinfo{person}{NVIDIA}.} \bibinfo{year}{2024}\natexlab{c}.
\newblock \bibinfo{title}{cuBLAS}.
\newblock
\newblock
\urldef\tempurl%
\url{https://docs.nvidia.com/cuda/cublas/}
\showURL{%
\tempurl}


\bibitem[NVIDIA(2024d)]%
        {cutlass}
\bibfield{author}{\bibinfo{person}{NVIDIA}.} \bibinfo{year}{2024}\natexlab{d}.
\newblock \bibinfo{title}{CUTLASS}.
\newblock
\newblock
\urldef\tempurl%
\url{https://nvidia.github.io/cutlass/}
\showURL{%
\tempurl}


\bibitem[NVIDIA(2024e)]%
        {hopperinline}
\bibfield{author}{\bibinfo{person}{NVIDIA}.} \bibinfo{year}{2024}\natexlab{e}.
\newblock \bibinfo{title}{Hopper Tuning Guide}.
\newblock
\newblock
\urldef\tempurl%
\url{https://docs.nvidia.com/cuda/hopper-tuning-guide/index.html#inline-compression}
\showURL{%
\tempurl}


\bibitem[NVIDIA(2024f)]%
        {turing}
\bibfield{author}{\bibinfo{person}{NVIDIA}.} \bibinfo{year}{2024}\natexlab{f}.
\newblock \bibinfo{title}{NVIDIA Turing Architecture In Depth}.
\newblock
\newblock
\urldef\tempurl%
\url{https://developer.nvidia.com/blog/nvidia-turing-architecture-in-depth/}
\showURL{%
\tempurl}


\bibitem[NVIDIA(2024g)]%
        {trtllm}
\bibfield{author}{\bibinfo{person}{NVIDIA}.} \bibinfo{year}{2024}\natexlab{g}.
\newblock \bibinfo{title}{TensorRT-LLM}.
\newblock
\newblock
\urldef\tempurl%
\url{https://nvidia.github.io/TensorRT-LLM/}
\showURL{%
\tempurl}


\bibitem[OpenAI et~al\mbox{.}(2024)]%
        {achiam2023gpt}
\bibfield{author}{\bibinfo{person}{OpenAI}, \bibinfo{person}{Josh Achiam}, \bibinfo{person}{Steven Adler}, \bibinfo{person}{Sandhini Agarwal}, \bibinfo{person}{Lama Ahmad}, \bibinfo{person}{Ilge Akkaya}, \bibinfo{person}{Florencia~Leoni Aleman}, \bibinfo{person}{Diogo Almeida}, \bibinfo{person}{Janko Altenschmidt}, \bibinfo{person}{Sam Altman}, \bibinfo{person}{Shyamal Anadkat}, \bibinfo{person}{Red Avila}, \bibinfo{person}{Igor Babuschkin}, \bibinfo{person}{Suchir Balaji}, \bibinfo{person}{Valerie Balcom}, \bibinfo{person}{Paul Baltescu}, \bibinfo{person}{Haiming Bao}, \bibinfo{person}{Mohammad Bavarian}, \bibinfo{person}{Jeff Belgum}, \bibinfo{person}{Irwan Bello}, \bibinfo{person}{Jake Berdine}, \bibinfo{person}{Gabriel Bernadett-Shapiro}, \bibinfo{person}{Christopher Berner}, \bibinfo{person}{Lenny Bogdonoff}, \bibinfo{person}{Oleg Boiko}, \bibinfo{person}{Madelaine Boyd}, \bibinfo{person}{Anna-Luisa Brakman}, \bibinfo{person}{Greg Brockman}, \bibinfo{person}{Tim Brooks}, \bibinfo{person}{Miles Brundage},
  \bibinfo{person}{Kevin Button}, \bibinfo{person}{Trevor Cai}, \bibinfo{person}{Rosie Campbell}, \bibinfo{person}{Andrew Cann}, \bibinfo{person}{Brittany Carey}, \bibinfo{person}{Chelsea Carlson}, \bibinfo{person}{Rory Carmichael}, \bibinfo{person}{Brooke Chan}, \bibinfo{person}{Che Chang}, \bibinfo{person}{Fotis Chantzis}, \bibinfo{person}{Derek Chen}, \bibinfo{person}{Sully Chen}, \bibinfo{person}{Ruby Chen}, \bibinfo{person}{Jason Chen}, \bibinfo{person}{Mark Chen}, \bibinfo{person}{Ben Chess}, \bibinfo{person}{Chester Cho}, \bibinfo{person}{Casey Chu}, \bibinfo{person}{Hyung~Won Chung}, \bibinfo{person}{Dave Cummings}, \bibinfo{person}{Jeremiah Currier}, \bibinfo{person}{Yunxing Dai}, \bibinfo{person}{Cory Decareaux}, \bibinfo{person}{Thomas Degry}, \bibinfo{person}{Noah Deutsch}, \bibinfo{person}{Damien Deville}, \bibinfo{person}{Arka Dhar}, \bibinfo{person}{David Dohan}, \bibinfo{person}{Steve Dowling}, \bibinfo{person}{Sheila Dunning}, \bibinfo{person}{Adrien Ecoffet}, \bibinfo{person}{Atty Eleti},
  \bibinfo{person}{Tyna Eloundou}, \bibinfo{person}{David Farhi}, \bibinfo{person}{Liam Fedus}, \bibinfo{person}{Niko Felix}, \bibinfo{person}{Simón~Posada Fishman}, \bibinfo{person}{Juston Forte}, \bibinfo{person}{Isabella Fulford}, \bibinfo{person}{Leo Gao}, \bibinfo{person}{Elie Georges}, \bibinfo{person}{Christian Gibson}, \bibinfo{person}{Vik Goel}, \bibinfo{person}{Tarun Gogineni}, \bibinfo{person}{Gabriel Goh}, \bibinfo{person}{Rapha Gontijo-Lopes}, \bibinfo{person}{Jonathan Gordon}, \bibinfo{person}{Morgan Grafstein}, \bibinfo{person}{Scott Gray}, \bibinfo{person}{Ryan Greene}, \bibinfo{person}{Joshua Gross}, \bibinfo{person}{Shixiang~Shane Gu}, \bibinfo{person}{Yufei Guo}, \bibinfo{person}{Chris Hallacy}, \bibinfo{person}{Jesse Han}, \bibinfo{person}{Jeff Harris}, \bibinfo{person}{Yuchen He}, \bibinfo{person}{Mike Heaton}, \bibinfo{person}{Johannes Heidecke}, \bibinfo{person}{Chris Hesse}, \bibinfo{person}{Alan Hickey}, \bibinfo{person}{Wade Hickey}, \bibinfo{person}{Peter Hoeschele},
  \bibinfo{person}{Brandon Houghton}, \bibinfo{person}{Kenny Hsu}, \bibinfo{person}{Shengli Hu}, \bibinfo{person}{Xin Hu}, \bibinfo{person}{Joost Huizinga}, \bibinfo{person}{Shantanu Jain}, \bibinfo{person}{Shawn Jain}, \bibinfo{person}{Joanne Jang}, \bibinfo{person}{Angela Jiang}, \bibinfo{person}{Roger Jiang}, \bibinfo{person}{Haozhun Jin}, \bibinfo{person}{Denny Jin}, \bibinfo{person}{Shino Jomoto}, \bibinfo{person}{Billie Jonn}, \bibinfo{person}{Heewoo Jun}, \bibinfo{person}{Tomer Kaftan}, \bibinfo{person}{Łukasz Kaiser}, \bibinfo{person}{Ali Kamali}, \bibinfo{person}{Ingmar Kanitscheider}, \bibinfo{person}{Nitish~Shirish Keskar}, \bibinfo{person}{Tabarak Khan}, \bibinfo{person}{Logan Kilpatrick}, \bibinfo{person}{Jong~Wook Kim}, \bibinfo{person}{Christina Kim}, \bibinfo{person}{Yongjik Kim}, \bibinfo{person}{Jan~Hendrik Kirchner}, \bibinfo{person}{Jamie Kiros}, \bibinfo{person}{Matt Knight}, \bibinfo{person}{Daniel Kokotajlo}, \bibinfo{person}{Łukasz Kondraciuk}, \bibinfo{person}{Andrew Kondrich},
  \bibinfo{person}{Aris Konstantinidis}, \bibinfo{person}{Kyle Kosic}, \bibinfo{person}{Gretchen Krueger}, \bibinfo{person}{Vishal Kuo}, \bibinfo{person}{Michael Lampe}, \bibinfo{person}{Ikai Lan}, \bibinfo{person}{Teddy Lee}, \bibinfo{person}{Jan Leike}, \bibinfo{person}{Jade Leung}, \bibinfo{person}{Daniel Levy}, \bibinfo{person}{Chak~Ming Li}, \bibinfo{person}{Rachel Lim}, \bibinfo{person}{Molly Lin}, \bibinfo{person}{Stephanie Lin}, \bibinfo{person}{Mateusz Litwin}, \bibinfo{person}{Theresa Lopez}, \bibinfo{person}{Ryan Lowe}, \bibinfo{person}{Patricia Lue}, \bibinfo{person}{Anna Makanju}, \bibinfo{person}{Kim Malfacini}, \bibinfo{person}{Sam Manning}, \bibinfo{person}{Todor Markov}, \bibinfo{person}{Yaniv Markovski}, \bibinfo{person}{Bianca Martin}, \bibinfo{person}{Katie Mayer}, \bibinfo{person}{Andrew Mayne}, \bibinfo{person}{Bob McGrew}, \bibinfo{person}{Scott~Mayer McKinney}, \bibinfo{person}{Christine McLeavey}, \bibinfo{person}{Paul McMillan}, \bibinfo{person}{Jake McNeil}, \bibinfo{person}{David
  Medina}, \bibinfo{person}{Aalok Mehta}, \bibinfo{person}{Jacob Menick}, \bibinfo{person}{Luke Metz}, \bibinfo{person}{Andrey Mishchenko}, \bibinfo{person}{Pamela Mishkin}, \bibinfo{person}{Vinnie Monaco}, \bibinfo{person}{Evan Morikawa}, \bibinfo{person}{Daniel Mossing}, \bibinfo{person}{Tong Mu}, \bibinfo{person}{Mira Murati}, \bibinfo{person}{Oleg Murk}, \bibinfo{person}{David Mély}, \bibinfo{person}{Ashvin Nair}, \bibinfo{person}{Reiichiro Nakano}, \bibinfo{person}{Rajeev Nayak}, \bibinfo{person}{Arvind Neelakantan}, \bibinfo{person}{Richard Ngo}, \bibinfo{person}{Hyeonwoo Noh}, \bibinfo{person}{Long Ouyang}, \bibinfo{person}{Cullen O'Keefe}, \bibinfo{person}{Jakub Pachocki}, \bibinfo{person}{Alex Paino}, \bibinfo{person}{Joe Palermo}, \bibinfo{person}{Ashley Pantuliano}, \bibinfo{person}{Giambattista Parascandolo}, \bibinfo{person}{Joel Parish}, \bibinfo{person}{Emy Parparita}, \bibinfo{person}{Alex Passos}, \bibinfo{person}{Mikhail Pavlov}, \bibinfo{person}{Andrew Peng}, \bibinfo{person}{Adam
  Perelman}, \bibinfo{person}{Filipe de Avila Belbute~Peres}, \bibinfo{person}{Michael Petrov}, \bibinfo{person}{Henrique~Ponde de Oliveira~Pinto}, \bibinfo{person}{Michael}, \bibinfo{person}{Pokorny}, \bibinfo{person}{Michelle Pokrass}, \bibinfo{person}{Vitchyr~H. Pong}, \bibinfo{person}{Tolly Powell}, \bibinfo{person}{Alethea Power}, \bibinfo{person}{Boris Power}, \bibinfo{person}{Elizabeth Proehl}, \bibinfo{person}{Raul Puri}, \bibinfo{person}{Alec Radford}, \bibinfo{person}{Jack Rae}, \bibinfo{person}{Aditya Ramesh}, \bibinfo{person}{Cameron Raymond}, \bibinfo{person}{Francis Real}, \bibinfo{person}{Kendra Rimbach}, \bibinfo{person}{Carl Ross}, \bibinfo{person}{Bob Rotsted}, \bibinfo{person}{Henri Roussez}, \bibinfo{person}{Nick Ryder}, \bibinfo{person}{Mario Saltarelli}, \bibinfo{person}{Ted Sanders}, \bibinfo{person}{Shibani Santurkar}, \bibinfo{person}{Girish Sastry}, \bibinfo{person}{Heather Schmidt}, \bibinfo{person}{David Schnurr}, \bibinfo{person}{John Schulman}, \bibinfo{person}{Daniel Selsam},
  \bibinfo{person}{Kyla Sheppard}, \bibinfo{person}{Toki Sherbakov}, \bibinfo{person}{Jessica Shieh}, \bibinfo{person}{Sarah Shoker}, \bibinfo{person}{Pranav Shyam}, \bibinfo{person}{Szymon Sidor}, \bibinfo{person}{Eric Sigler}, \bibinfo{person}{Maddie Simens}, \bibinfo{person}{Jordan Sitkin}, \bibinfo{person}{Katarina Slama}, \bibinfo{person}{Ian Sohl}, \bibinfo{person}{Benjamin Sokolowsky}, \bibinfo{person}{Yang Song}, \bibinfo{person}{Natalie Staudacher}, \bibinfo{person}{Felipe~Petroski Such}, \bibinfo{person}{Natalie Summers}, \bibinfo{person}{Ilya Sutskever}, \bibinfo{person}{Jie Tang}, \bibinfo{person}{Nikolas Tezak}, \bibinfo{person}{Madeleine~B. Thompson}, \bibinfo{person}{Phil Tillet}, \bibinfo{person}{Amin Tootoonchian}, \bibinfo{person}{Elizabeth Tseng}, \bibinfo{person}{Preston Tuggle}, \bibinfo{person}{Nick Turley}, \bibinfo{person}{Jerry Tworek}, \bibinfo{person}{Juan Felipe~Cerón Uribe}, \bibinfo{person}{Andrea Vallone}, \bibinfo{person}{Arun Vijayvergiya}, \bibinfo{person}{Chelsea Voss},
  \bibinfo{person}{Carroll Wainwright}, \bibinfo{person}{Justin~Jay Wang}, \bibinfo{person}{Alvin Wang}, \bibinfo{person}{Ben Wang}, \bibinfo{person}{Jonathan Ward}, \bibinfo{person}{Jason Wei}, \bibinfo{person}{CJ Weinmann}, \bibinfo{person}{Akila Welihinda}, \bibinfo{person}{Peter Welinder}, \bibinfo{person}{Jiayi Weng}, \bibinfo{person}{Lilian Weng}, \bibinfo{person}{Matt Wiethoff}, \bibinfo{person}{Dave Willner}, \bibinfo{person}{Clemens Winter}, \bibinfo{person}{Samuel Wolrich}, \bibinfo{person}{Hannah Wong}, \bibinfo{person}{Lauren Workman}, \bibinfo{person}{Sherwin Wu}, \bibinfo{person}{Jeff Wu}, \bibinfo{person}{Michael Wu}, \bibinfo{person}{Kai Xiao}, \bibinfo{person}{Tao Xu}, \bibinfo{person}{Sarah Yoo}, \bibinfo{person}{Kevin Yu}, \bibinfo{person}{Qiming Yuan}, \bibinfo{person}{Wojciech Zaremba}, \bibinfo{person}{Rowan Zellers}, \bibinfo{person}{Chong Zhang}, \bibinfo{person}{Marvin Zhang}, \bibinfo{person}{Shengjia Zhao}, \bibinfo{person}{Tianhao Zheng}, \bibinfo{person}{Juntang Zhuang},
  \bibinfo{person}{William Zhuk}, {and} \bibinfo{person}{Barret Zoph}.} \bibinfo{year}{2024}\natexlab{}.
\newblock \bibinfo{title}{GPT-4 Technical Report}.
\newblock
\newblock
\showeprint[arxiv]{2303.08774}~[cs.CL]
\urldef\tempurl%
\url{https://arxiv.org/abs/2303.08774}
\showURL{%
\tempurl}


\bibitem[Patel et~al\mbox{.}(2024)]%
        {patel2024splitwiseefficientgenerativellm}
\bibfield{author}{\bibinfo{person}{Pratyush Patel}, \bibinfo{person}{Esha Choukse}, \bibinfo{person}{Chaojie Zhang}, \bibinfo{person}{Aashaka Shah}, \bibinfo{person}{Íñigo Goiri}, \bibinfo{person}{Saeed Maleki}, {and} \bibinfo{person}{Ricardo Bianchini}.} \bibinfo{year}{2024}\natexlab{}.
\newblock \bibinfo{title}{Splitwise: Efficient generative LLM inference using phase splitting}.
\newblock
\newblock
\showeprint[arxiv]{2311.18677}~[cs.AR]
\urldef\tempurl%
\url{https://arxiv.org/abs/2311.18677}
\showURL{%
\tempurl}


\bibitem[Pekhimenko et~al\mbox{.}(2012)]%
        {pekhimenko2012base}
\bibfield{author}{\bibinfo{person}{Gennady Pekhimenko}, \bibinfo{person}{Vivek Seshadri}, \bibinfo{person}{Onur Mutlu}, \bibinfo{person}{Phillip~B. Gibbons}, \bibinfo{person}{Michael~A. Kozuch}, {and} \bibinfo{person}{Todd~C. Mowry}.} \bibinfo{year}{2012}\natexlab{}.
\newblock \showarticletitle{Base-delta-immediate compression: practical data compression for on-chip caches}. In \bibinfo{booktitle}{\emph{Proceedings of the 21st International Conference on Parallel Architectures and Compilation Techniques}} (Minneapolis, Minnesota, USA) \emph{(\bibinfo{series}{PACT '12})}. \bibinfo{publisher}{Association for Computing Machinery}, \bibinfo{address}{New York, NY, USA}, \bibinfo{pages}{377–388}.
\newblock
\showISBNx{9781450311823}
\urldef\tempurl%
\url{https://doi.org/10.1145/2370816.2370870}
\showDOI{\tempurl}


\bibitem[Pyzer-Knapp et~al\mbox{.}(2022)]%
        {hpc1}
\bibfield{author}{\bibinfo{person}{Edward~O Pyzer-Knapp}, \bibinfo{person}{Jed~W Pitera}, \bibinfo{person}{Peter~WJ Staar}, \bibinfo{person}{Seiji Takeda}, \bibinfo{person}{Teodoro Laino}, \bibinfo{person}{Daniel~P Sanders}, \bibinfo{person}{James Sexton}, \bibinfo{person}{John~R Smith}, {and} \bibinfo{person}{Alessandro Curioni}.} \bibinfo{year}{2022}\natexlab{}.
\newblock \showarticletitle{Accelerating materials discovery using artificial intelligence, high performance computing and robotics}.
\newblock \bibinfo{journal}{\emph{npj Computational Materials}} \bibinfo{volume}{8}, \bibinfo{number}{1} (\bibinfo{year}{2022}), \bibinfo{pages}{84}.
\newblock


\bibitem[Sakaguchi et~al\mbox{.}(2021)]%
        {winogrande}
\bibfield{author}{\bibinfo{person}{Keisuke Sakaguchi}, \bibinfo{person}{Ronan~Le Bras}, \bibinfo{person}{Chandra Bhagavatula}, {and} \bibinfo{person}{Yejin Choi}.} \bibinfo{year}{2021}\natexlab{}.
\newblock \showarticletitle{Winogrande: An adversarial winograd schema challenge at scale}.
\newblock \bibinfo{journal}{\emph{Commun. ACM}} \bibinfo{volume}{64}, \bibinfo{number}{9} (\bibinfo{year}{2021}), \bibinfo{pages}{99--106}.
\newblock


\bibitem[Shoeybi et~al\mbox{.}(2020)]%
        {shoeybi2019megatron}
\bibfield{author}{\bibinfo{person}{Mohammad Shoeybi}, \bibinfo{person}{Mostofa Patwary}, \bibinfo{person}{Raul Puri}, \bibinfo{person}{Patrick LeGresley}, \bibinfo{person}{Jared Casper}, {and} \bibinfo{person}{Bryan Catanzaro}.} \bibinfo{year}{2020}\natexlab{}.
\newblock \bibinfo{title}{Megatron-LM: Training Multi-Billion Parameter Language Models Using Model Parallelism}.
\newblock
\newblock
\showeprint[arxiv]{1909.08053}~[cs.CL]
\urldef\tempurl%
\url{https://arxiv.org/abs/1909.08053}
\showURL{%
\tempurl}


\bibitem[Sterling et~al\mbox{.}(2017)]%
        {hpc2}
\bibfield{author}{\bibinfo{person}{Thomas Sterling}, \bibinfo{person}{Maciej Brodowicz}, {and} \bibinfo{person}{Matthew Anderson}.} \bibinfo{year}{2017}\natexlab{}.
\newblock \bibinfo{booktitle}{\emph{High performance computing: modern systems and practices}}.
\newblock \bibinfo{publisher}{Morgan Kaufmann}.
\newblock


\bibitem[Touvron et~al\mbox{.}(2023a)]%
        {touvron2023llama}
\bibfield{author}{\bibinfo{person}{Hugo Touvron}, \bibinfo{person}{Thibaut Lavril}, \bibinfo{person}{Gautier Izacard}, \bibinfo{person}{Xavier Martinet}, \bibinfo{person}{Marie-Anne Lachaux}, \bibinfo{person}{Timothée Lacroix}, \bibinfo{person}{Baptiste Rozière}, \bibinfo{person}{Naman Goyal}, \bibinfo{person}{Eric Hambro}, \bibinfo{person}{Faisal Azhar}, \bibinfo{person}{Aurelien Rodriguez}, \bibinfo{person}{Armand Joulin}, \bibinfo{person}{Edouard Grave}, {and} \bibinfo{person}{Guillaume Lample}.} \bibinfo{year}{2023}\natexlab{a}.
\newblock \bibinfo{title}{LLaMA: Open and Efficient Foundation Language Models}.
\newblock
\newblock
\showeprint[arxiv]{2302.13971}~[cs.CL]
\urldef\tempurl%
\url{https://arxiv.org/abs/2302.13971}
\showURL{%
\tempurl}


\bibitem[Touvron et~al\mbox{.}(2023b)]%
        {llama2}
\bibfield{author}{\bibinfo{person}{Hugo Touvron}, \bibinfo{person}{Louis Martin}, \bibinfo{person}{Kevin Stone}, \bibinfo{person}{Peter Albert}, \bibinfo{person}{Amjad Almahairi}, \bibinfo{person}{Yasmine Babaei}, \bibinfo{person}{Nikolay Bashlykov}, \bibinfo{person}{Soumya Batra}, \bibinfo{person}{Prajjwal Bhargava}, \bibinfo{person}{Shruti Bhosale}, \bibinfo{person}{Dan Bikel}, \bibinfo{person}{Lukas Blecher}, \bibinfo{person}{Cristian~Canton Ferrer}, \bibinfo{person}{Moya Chen}, \bibinfo{person}{Guillem Cucurull}, \bibinfo{person}{David Esiobu}, \bibinfo{person}{Jude Fernandes}, \bibinfo{person}{Jeremy Fu}, \bibinfo{person}{Wenyin Fu}, \bibinfo{person}{Brian Fuller}, \bibinfo{person}{Cynthia Gao}, \bibinfo{person}{Vedanuj Goswami}, \bibinfo{person}{Naman Goyal}, \bibinfo{person}{Anthony Hartshorn}, \bibinfo{person}{Saghar Hosseini}, \bibinfo{person}{Rui Hou}, \bibinfo{person}{Hakan Inan}, \bibinfo{person}{Marcin Kardas}, \bibinfo{person}{Viktor Kerkez}, \bibinfo{person}{Madian Khabsa},
  \bibinfo{person}{Isabel Kloumann}, \bibinfo{person}{Artem Korenev}, \bibinfo{person}{Punit~Singh Koura}, \bibinfo{person}{Marie-Anne Lachaux}, \bibinfo{person}{Thibaut Lavril}, \bibinfo{person}{Jenya Lee}, \bibinfo{person}{Diana Liskovich}, \bibinfo{person}{Yinghai Lu}, \bibinfo{person}{Yuning Mao}, \bibinfo{person}{Xavier Martinet}, \bibinfo{person}{Todor Mihaylov}, \bibinfo{person}{Pushkar Mishra}, \bibinfo{person}{Igor Molybog}, \bibinfo{person}{Yixin Nie}, \bibinfo{person}{Andrew Poulton}, \bibinfo{person}{Jeremy Reizenstein}, \bibinfo{person}{Rashi Rungta}, \bibinfo{person}{Kalyan Saladi}, \bibinfo{person}{Alan Schelten}, \bibinfo{person}{Ruan Silva}, \bibinfo{person}{Eric~Michael Smith}, \bibinfo{person}{Ranjan Subramanian}, \bibinfo{person}{Xiaoqing~Ellen Tan}, \bibinfo{person}{Binh Tang}, \bibinfo{person}{Ross Taylor}, \bibinfo{person}{Adina Williams}, \bibinfo{person}{Jian~Xiang Kuan}, \bibinfo{person}{Puxin Xu}, \bibinfo{person}{Zheng Yan}, \bibinfo{person}{Iliyan Zarov}, \bibinfo{person}{Yuchen
  Zhang}, \bibinfo{person}{Angela Fan}, \bibinfo{person}{Melanie Kambadur}, \bibinfo{person}{Sharan Narang}, \bibinfo{person}{Aurelien Rodriguez}, \bibinfo{person}{Robert Stojnic}, \bibinfo{person}{Sergey Edunov}, {and} \bibinfo{person}{Thomas Scialom}.} \bibinfo{year}{2023}\natexlab{b}.
\newblock \bibinfo{title}{Llama 2: Open Foundation and Fine-Tuned Chat Models}.
\newblock
\newblock
\showeprint[arxiv]{2307.09288}~[cs.CL]
\urldef\tempurl%
\url{https://arxiv.org/abs/2307.09288}
\showURL{%
\tempurl}


\bibitem[Vaswani et~al\mbox{.}(2023)]%
        {vaswani2023attentionneed}
\bibfield{author}{\bibinfo{person}{Ashish Vaswani}, \bibinfo{person}{Noam Shazeer}, \bibinfo{person}{Niki Parmar}, \bibinfo{person}{Jakob Uszkoreit}, \bibinfo{person}{Llion Jones}, \bibinfo{person}{Aidan~N. Gomez}, \bibinfo{person}{Lukasz Kaiser}, {and} \bibinfo{person}{Illia Polosukhin}.} \bibinfo{year}{2023}\natexlab{}.
\newblock \bibinfo{title}{Attention Is All You Need}.
\newblock
\newblock
\showeprint[arxiv]{1706.03762}~[cs.CL]
\urldef\tempurl%
\url{https://arxiv.org/abs/1706.03762}
\showURL{%
\tempurl}


\bibitem[Villa et~al\mbox{.}(2019)]%
        {nvbit}
\bibfield{author}{\bibinfo{person}{Oreste Villa}, \bibinfo{person}{Mark Stephenson}, \bibinfo{person}{David Nellans}, {and} \bibinfo{person}{Stephen~W. Keckler}.} \bibinfo{year}{2019}\natexlab{}.
\newblock \showarticletitle{NVBit: A Dynamic Binary Instrumentation Framework for NVIDIA GPUs}. In \bibinfo{booktitle}{\emph{Proceedings of the 52nd Annual IEEE/ACM International Symposium on Microarchitecture}} (Columbus, OH, USA) \emph{(\bibinfo{series}{MICRO '52})}. \bibinfo{publisher}{Association for Computing Machinery}, \bibinfo{address}{New York, NY, USA}, \bibinfo{pages}{372–383}.
\newblock
\showISBNx{9781450369381}
\urldef\tempurl%
\url{https://doi.org/10.1145/3352460.3358307}
\showDOI{\tempurl}


\bibitem[Wallace(1992)]%
        {wallace1992jpeg}
\bibfield{author}{\bibinfo{person}{Gregory~K Wallace}.} \bibinfo{year}{1992}\natexlab{}.
\newblock \showarticletitle{The {JPEG} still picture compression standard}.
\newblock \bibinfo{journal}{\emph{IEEE Transactions on Consumer Electronics}} \bibinfo{volume}{38}, \bibinfo{number}{1} (\bibinfo{year}{1992}), \bibinfo{pages}{xviii--xxxiv}.
\newblock


\bibitem[Wei et~al\mbox{.}(2025)]%
        {wei2025prosperity}
\bibfield{author}{\bibinfo{person}{Chiyue Wei}, \bibinfo{person}{Cong Guo}, \bibinfo{person}{Feng Cheng}, \bibinfo{person}{Shiyu Li}, \bibinfo{person}{Hao~Frank Yang}, \bibinfo{person}{Hai~Helen Li}, {and} \bibinfo{person}{Yiran Chen}.} \bibinfo{year}{2025}\natexlab{}.
\newblock \showarticletitle{Prosperity: Accelerating Spiking Neural Networks via Product Sparsity}. In \bibinfo{booktitle}{\emph{2025 IEEE International Symposium on High Performance Computer Architecture (HPCA)}}. IEEE, \bibinfo{pages}{806--820}.
\newblock


\bibitem[Xiao et~al\mbox{.}(2024)]%
        {smoothquant}
\bibfield{author}{\bibinfo{person}{Guangxuan Xiao}, \bibinfo{person}{Ji Lin}, \bibinfo{person}{Mickael Seznec}, \bibinfo{person}{Hao Wu}, \bibinfo{person}{Julien Demouth}, {and} \bibinfo{person}{Song Han}.} \bibinfo{year}{2024}\natexlab{}.
\newblock \bibinfo{title}{SmoothQuant: Accurate and Efficient Post-Training Quantization for Large Language Models}.
\newblock
\newblock
\showeprint[arxiv]{2211.10438}~[cs.CL]
\urldef\tempurl%
\url{https://arxiv.org/abs/2211.10438}
\showURL{%
\tempurl}


\bibitem[Yao et~al\mbox{.}(2023)]%
        {llm_solve}
\bibfield{author}{\bibinfo{person}{Shunyu Yao}, \bibinfo{person}{Dian Yu}, \bibinfo{person}{Jeffrey Zhao}, \bibinfo{person}{Izhak Shafran}, \bibinfo{person}{Thomas~L. Griffiths}, \bibinfo{person}{Yuan Cao}, {and} \bibinfo{person}{Karthik Narasimhan}.} \bibinfo{year}{2023}\natexlab{}.
\newblock \bibinfo{title}{Tree of Thoughts: Deliberate Problem Solving with Large Language Models}.
\newblock
\newblock
\showeprint[arxiv]{2305.10601}~[cs.CL]
\urldef\tempurl%
\url{https://arxiv.org/abs/2305.10601}
\showURL{%
\tempurl}


\bibitem[Yuan et~al\mbox{.}(2024)]%
        {yuan2024llm}
\bibfield{author}{\bibinfo{person}{Zhihang Yuan}, \bibinfo{person}{Yuzhang Shang}, \bibinfo{person}{Yang Zhou}, \bibinfo{person}{Zhen Dong}, \bibinfo{person}{Zhe Zhou}, \bibinfo{person}{Chenhao Xue}, \bibinfo{person}{Bingzhe Wu}, \bibinfo{person}{Zhikai Li}, \bibinfo{person}{Qingyi Gu}, \bibinfo{person}{Yong~Jae Lee}, \bibinfo{person}{Yan Yan}, \bibinfo{person}{Beidi Chen}, \bibinfo{person}{Guangyu Sun}, {and} \bibinfo{person}{Kurt Keutzer}.} \bibinfo{year}{2024}\natexlab{}.
\newblock \bibinfo{title}{LLM Inference Unveiled: Survey and Roofline Model Insights}.
\newblock
\newblock
\showeprint[arxiv]{2402.16363}~[cs.CL]
\urldef\tempurl%
\url{https://arxiv.org/abs/2402.16363}
\showURL{%
\tempurl}


\bibitem[Zellers et~al\mbox{.}(2019)]%
        {hellaswag}
\bibfield{author}{\bibinfo{person}{Rowan Zellers}, \bibinfo{person}{Ari Holtzman}, \bibinfo{person}{Yonatan Bisk}, \bibinfo{person}{Ali Farhadi}, {and} \bibinfo{person}{Yejin Choi}.} \bibinfo{year}{2019}\natexlab{}.
\newblock \bibinfo{title}{HellaSwag: Can a Machine Really Finish Your Sentence?}
\newblock
\newblock
\showeprint[arxiv]{1905.07830}~[cs.CL]
\urldef\tempurl%
\url{https://arxiv.org/abs/1905.07830}
\showURL{%
\tempurl}


\bibitem[Zhang et~al\mbox{.}(2024b)]%
        {zhang2024dstc}
\bibfield{author}{\bibinfo{person}{Chen Zhang}, \bibinfo{person}{Yang Wang}, \bibinfo{person}{Zhiqiang Xie}, \bibinfo{person}{Cong Guo}, \bibinfo{person}{Yunxin Liu}, \bibinfo{person}{Jingwen Leng}, \bibinfo{person}{Guangyu Sun}, \bibinfo{person}{Zhigang Ji}, \bibinfo{person}{Runsheng Wang}, \bibinfo{person}{Yuan Xie}, {et~al\mbox{.}}} \bibinfo{year}{2024}\natexlab{b}.
\newblock \showarticletitle{DSTC: Dual-Side Sparsity Tensor Core for DNNs Acceleration on Modern GPU Architectures}.
\newblock \bibinfo{journal}{\emph{IEEE Trans. Comput.}} (\bibinfo{year}{2024}).
\newblock


\bibitem[Zhang et~al\mbox{.}(2024a)]%
        {hpc3}
\bibfield{author}{\bibinfo{person}{Junyao Zhang}, \bibinfo{person}{Hanrui Wang}, \bibinfo{person}{Qi Ding}, \bibinfo{person}{Jiaqi Gu}, \bibinfo{person}{Reouven Assouly}, \bibinfo{person}{William~D Oliver}, \bibinfo{person}{Song Han}, \bibinfo{person}{Kenneth~R Brown}, \bibinfo{person}{Hai Li}, \bibinfo{person}{Yiran Chen}, {et~al\mbox{.}}} \bibinfo{year}{2024}\natexlab{a}.
\newblock \showarticletitle{Qplacer: Frequency-aware component placement for superconducting quantum computers}.
\newblock \bibinfo{journal}{\emph{arXiv preprint arXiv:2401.17450}} (\bibinfo{year}{2024}).
\newblock


\end{thebibliography}
%%%%%%%%%%%%%%%%%%%%%%%%%%%%%%%%%%%%

\end{document}